\PassOptionsToPackage{unicode}{hyperref}
\PassOptionsToPackage{hyphens}{url}
\PassOptionsToPackage{dvipsnames,svgnames,x11names}{xcolor}
\documentclass[
  letterpaper,
  DIV=11,
  numbers=noendperiod]{scrartcl}

\usepackage{amsmath,amssymb}
\usepackage{iftex}
\ifPDFTeX
  \usepackage[T1]{fontenc}
  \usepackage[utf8]{inputenc}
  \usepackage{textcomp} 
\else 
  \usepackage{unicode-math}
  \defaultfontfeatures{Scale=MatchLowercase}
  \defaultfontfeatures[\rmfamily]{Ligatures=TeX,Scale=1}
\fi
\usepackage{lmodern}
\ifPDFTeX\else  
\fi
\IfFileExists{upquote.sty}{\usepackage{upquote}}{}
\IfFileExists{microtype.sty}{
  \usepackage[]{microtype}
  \UseMicrotypeSet[protrusion]{basicmath} 
}{}
\makeatletter
\@ifundefined{KOMAClassName}{
  \IfFileExists{parskip.sty}{%
    \usepackage{parskip}
  }{
    \setlength{\parindent}{0pt}
    \setlength{\parskip}{6pt plus 2pt minus 1pt}}
}{
  \KOMAoptions{parskip=half}}
\makeatother
\usepackage{xcolor}
\makeatletter
\ifx\paragraph\undefined\else
  \let\oldparagraph\paragraph
  \renewcommand{\paragraph}{
    \@ifstar
      \xxxParagraphStar
      \xxxParagraphNoStar
  }
  \newcommand{\xxxParagraphStar}[1]{\oldparagraph*{#1}\mbox{}}
  \newcommand{\xxxParagraphNoStar}[1]{\oldparagraph{#1}\mbox{}}
\fi
\ifx\subparagraph\undefined\else
  \let\oldsubparagraph\subparagraph
  \renewcommand{\subparagraph}{
    \@ifstar
      \xxxSubParagraphStar
      \xxxSubParagraphNoStar
  }
  \newcommand{\xxxSubParagraphStar}[1]{\oldsubparagraph*{#1}\mbox{}}
  \newcommand{\xxxSubParagraphNoStar}[1]{\oldsubparagraph{#1}\mbox{}}
\fi
\makeatother

\providecommand{\tightlist}{%
  \setlength{\itemsep}{0pt}\setlength{\parskip}{0pt}}\usepackage{longtable,booktabs,array}
\usepackage{calc} 
\usepackage{etoolbox}
\makeatletter
\patchcmd\longtable{\par}{\if@noskipsec\mbox{}\fi\par}{}{}
\makeatother
\IfFileExists{footnotehyper.sty}{\usepackage{footnotehyper}}{\usepackage{footnote}}
\makesavenoteenv{longtable}
\usepackage{graphicx}
\makeatletter
\newsavebox\pandoc@box
\newcommand*\pandocbounded[1]{
  \sbox\pandoc@box{#1}%
  \Gscale@div\@tempa{\textheight}{\dimexpr\ht\pandoc@box+\dp\pandoc@box\relax}%
  \Gscale@div\@tempb{\linewidth}{\wd\pandoc@box}%
  \ifdim\@tempb\p@<\@tempa\p@\let\@tempa\@tempb\fi
  \ifdim\@tempa\p@<\p@\scalebox{\@tempa}{\usebox\pandoc@box}%
  \else\usebox{\pandoc@box}%
  \fi%
}
\def\fps@figure{htbp}
\makeatother
\NewDocumentCommand\citeproctext{}{}

\makeatletter
 \let\@cite@ofmt\@firstofone
 \def\@biblabel#1{}
 \def\@cite#1#2{{#1\if@tempswa , #2\fi}}
\makeatother
\newlength{\cslhangindent}
\newlength{\csllabelwidth}
\newenvironment{CSLReferences}[2] 
 {\begin{list}{}{%
  \setlength{\itemindent}{0pt}
  \setlength{\leftmargin}{0pt}
  \setlength{\parsep}{0pt}
  \ifodd #1
   \setlength{\leftmargin}{\cslhangindent}
   \setlength{\itemindent}{-1\cslhangindent}
  \fi
  \setlength{\itemsep}{#2\baselineskip}}}
 {\end{list}}
\usepackage{calc}

\usepackage{booktabs}
\usepackage{longtable}
\usepackage{array}
\usepackage{multirow}
\usepackage{wrapfig}
\usepackage{float}
\usepackage{colortbl}
\usepackage{pdflscape}
\usepackage{tabu}
\usepackage{threeparttable}
\usepackage{threeparttablex}
\usepackage[normalem]{ulem}
\usepackage{makecell}
\usepackage{xcolor}
\KOMAoption{captions}{tableheading}
\makeatletter
\@ifpackageloaded{caption}{}{\usepackage{caption}}
\AtBeginDocument{%
\ifdefined\contentsname
  \renewcommand*\contentsname{Table of contents}
\else
  \newcommand\contentsname{Table of contents}
\fi
\ifdefined\listfigurename
  \renewcommand*\listfigurename{List of Figures}
\else
  \newcommand\listfigurename{List of Figures}
\fi
\ifdefined\listtablename
  \renewcommand*\listtablename{List of Tables}
\else
  \newcommand\listtablename{List of Tables}
\fi
\ifdefined\figurename
  \renewcommand*\figurename{Figure}
\else
  \newcommand\figurename{Figure}
\fi
\ifdefined\tablename
  \renewcommand*\tablename{Table}
\else
  \newcommand\tablename{Table}
\fi
}
\@ifpackageloaded{float}{}{\usepackage{float}}
\floatstyle{ruled}
\@ifundefined{c@chapter}{\newfloat{codelisting}{h}{lop}}{\newfloat{codelisting}{h}{lop}[chapter]}
\floatname{codelisting}{Listing}

\makeatother
\makeatletter
\@ifpackageloaded{caption}{}{\usepackage{caption}}
\@ifpackageloaded{subcaption}{}{\usepackage{subcaption}}
\makeatother

\usepackage{bookmark}

\IfFileExists{xurl.sty}{\usepackage{xurl}}{} 
\hypersetup{
  pdftitle={Prompting the Professoriate: A Qualitative Study of Instructor Perspectives on LLMs in Data Science Education},
  pdfauthor={Ana Elisa Lopez-Miranda; Tiffany Timbers; Rohan Alexander},
  colorlinks=true,
  linkcolor={blue},
  filecolor={Maroon},
  citecolor={Blue},
  urlcolor={Blue},
  pdfcreator={LaTeX via pandoc}}

\title{Prompting the Professoriate: A Qualitative Study of Instructor
Perspectives on LLMs in Data Science Education\thanks{Corresponding
author:
\href{mailto:rohan.alexander@utoronto.ca}{\nolinkurl{rohan.alexander@utoronto.ca}}.
CRediT contributions: \textbf{Ana Elisa Lopez-Miranda:}
Conceptualization; Data curation; Formal analysis; Investigation;
Methodology; Software; Validation; Writing -- original draft; Writing --
review \& editing. \textbf{Tiffany Timbers:} Conceptualization; Data
curation; Formal analysis; Investigation; Methodology; Software;
Supervision; Validation; Visualization; Writing -- original draft;
Writing -- review \& editing. \textbf{Rohan Alexander:}
Conceptualization; Data curation; Formal analysis; Funding acquisition;
Investigation; Project administration; Software; Supervision;
Validation; Writing -- original draft; Writing -- review \& editing.}}
\author{Ana Elisa Lopez-Miranda \and Tiffany Timbers \and Rohan
Alexander}
\date{September 14, 2025}

\begin{document}
\maketitle
\begin{abstract}
Large Language Models (LLMs) have shifted in just a few years from
novelty to ubiquity, raising fundamental questions for data science
education. Tasks once used to teach coding, writing, and problem-solving
can now be completed by LLMs, forcing educators to reconsider both
pedagogy and assessment. To understand how instructors are adapting, we
conducted semi-structured interviews with 42 instructors from 33
institutions in 10 countries in June and July 2025. Our qualitative
analysis reveals a pragmatic mix of optimism and concern. Many
respondents view LLMs as inevitable classroom tools---comparable to
calculators or Wikipedia---while others worry about de-skilling,
misplaced confidence, and uneven integration across institutions. Around
58 per cent have already introduced demonstrations, guided activities,
or make extensive use of LLMs in their courses, though most expect
change to remain slow and uneven. That said, 31 per cent have not used
LLMs to teach students and do not plan to. We highlight some
instructional innovations, including AI-aware assessments, reflective
use of LLMs as tutors, and course-specific chatbots. By sharing these
perspectives, we aim to help data science educators adapt collectively
to ensure curricula keep pace with technological change.
\end{abstract}


\section{Introduction}\label{sec-intro}

Data science is the study, development, and practice of using
reproducible and transparent processes to generate insight from data
(Berman et al. 2018; Wing 2019; Irizarry 2020; Timbers, Campbell, and
Lee 2022). Question and problem formulation, along with solution design
are perhaps the most central and important skills needed for generating
insight from data, however many other tasks are critical for data
science studies and applications. Examples include writing computer code
to create data visualizations, carry out analysis, train machine
learning and/or AI models, and deploy data science applications, as well
as writing narrative prose and creating effective presentations to
communicate insights and findings (Donoho 2017; Irizarry 2020). As a
consequence, data science courses and programs involve teaching students
the theory, concepts, and practical skills needed to perform these
diverse tasks.

With roots in statistics and computer science, data science educators
use many of the pedagogies and assessment strategies used in statistics
and computer science education (GAISE College Report ASA Revision
Committee 2016; Zendler and Klaudt 2015; Fincher and Robins 2019).
Having evolved into a distinct field of its own, with great emphasis on
collaboration (often within an interdisciplinary setting), data science
educators frequently tailor pedagogies and assessments that emphasize
real-world (and often messy) data, case studies, authentic data science
technologies, reproducibility, projects and teamwork (Memarian and
Doleck 2024).

Teaching has always required adaptation to new tools, whether
calculators, the internet, or Wikipedia. Each innovation reshaped what
students needed to know and how they came to learn it. Large Language
Models (LLMs) --- a type of artificial intelligence trained on vast
amounts of text to understand and generate human-like language, with
ChatGPT being the most well-known example --- likely represent the most
disruptive change yet. LLMs have become ubiquitous since the public
release of OpenAI's ChatGPT in November 2022. They can write
high-quality code and prose (Touvron et al. 2023; Hou et al. 2024; Liang
et al. 2024; Dingle and Kruliš 2024; Moradi Dakhel et al. 2023; Da
Silva, Samhi, and Khomh 2025), and perform data analysis (Cheng, Li, and
Bing 2023; Miranda and Campelo 2024; Maojun et al. 2025). Their training
on internet-scale data means they draw directly from the kinds of
resources that form the backbone of data science education, including
blogs, open educational resources (e.g., data science textbooks and
assignments), popular code repositories (e.g., GitHub and Bitbucket),
and public question and answer platforms (e.g., Stack Overflow and Stack
Exchange). As a result, LLMs can act as teaching assistants, as well as
help with curriculum design and pedagogical innovation (Ellis and Slade
2023; Tu et al. 2024). A less desirable, but also possible use of LLMs,
is as a substitute for student work. LLMs can answer many of the
questions typically posed to statistics and data science students in
their course assessments (Ellis and Slade 2023; Tu et al. 2024; Dingle
and Kruliš 2024; Maojun et al. 2025). This has negative implications for
both student learning and academic integrity.

For instructors, the challenge is twofold. On the one hand, LLMs promise
efficiency, personalization, and expanded access to expertise. On the
other, they raise serious concerns about ethics, bias, academic
integrity, and the erosion of core skills. The pace of adoption leaves
many educators disoriented: unsure how to adapt their courses, isolated
from peers facing similar questions, and still grappling with
pandemic-related pressures.

We imagine that many data science educators have made some promising
adjustments to their teaching in response to LLMs, such as encouraging
or restricting LLM use in assessments, adding new content to our
curriculum (e.g., transformers and LLMs), and changing approaches to
pedagogy. However, in a single instructor's teaching practices, these
changes can feel small and incremental. We suspect we are not alone in
this feeling. This paper is grounded in the hypothesis that many
educators are navigating similar uncertainties. By sharing what
instructors across institutions are thinking, feeling and doing, we hope
to foster a sense of community, spark and cross-pollinate new ideas
across institutions, and support one another in reimagining our teaching
during this pivotal moment.

To accomplish this, we conducted semi-structured interviews with 42
instructors teaching data science, and closely related, courses from 33
unique institutions in 10 countries, between 9 June and 9 July 2025.
During those interviews, instructors were asked about their thoughts and
approaches to LLMs in education. These interviews were conducted by one
of the authors and focused on three aspects:

\begin{enumerate}
\def\labelenumi{\arabic{enumi})}
\tightlist
\item
  instructor and course context;
\item
  LLM adoption and classroom practice changes;
\item
  assessment, learning impact, and academic integrity.
\end{enumerate}

From our qualitative analysis of the interview transcripts, we find
mixed views about LLMs. On the one hand, many respondents accept LLMs as
something that will be around and has changed what it means to do data
science. They recognize beneficial aspects such as the potential for
efficiency improvements, especially with coding, and developing initial
ideas. However many respondents also talked about their concerns about
ethics, environmental impact, bias, educational disruption, and the loss
of critical thinking. Respondents were worried about the potential for
uncritical adoption of LLMs and the potential for cognitive offloading
that could undermine skill development. They were also concerned about
the potential for increased skills gaps, and had considerable concerns
about whether academia would be able to adapt quickly enough to be able
to keep up. We also highlight some of the innovative, interesting, and
exciting ways data science instructors are adapting their teaching to
this new landscape.

Like Wikipedia, the internet, the calculator and many other inventions
before it, LLMs force us to consider what students need to know and how
they should come to learn it. Faculty do not know how the job market
will evolve and sometimes lack comfort with LLMs themselves. There is
also limited time and incentive for them to do the work to change.
Sharing these interviews, and our findings from analyzing them, can help
data science educators move in a more coordinated way which in turn can
help ensure pedagogies and curricula keep pace with changing
expectations of industry and students.

The remainder of this paper is structured as follows.
Section~\ref{sec-background} provides context to this paper,
Section~\ref{sec-design} details our study design,
Section~\ref{sec-results} specifies our results,
Section~\ref{sec-discussion} discusses some of the implications of our
results, as well as some limitations. Appendices provide details about
the initial engagement email, our interview guide, the steps we took to
anonymize transcripts, and our codebook. Anonymized transcripts, and our
coding of them, are available in an accompanying repository.

\section{Background}\label{sec-background}

Since the public release of OpenAI's ChatGPT in November 2022, public
awareness of the potential uses of AI, more specifically LLMs, has
substantially increased. While ``AI'' is a term in general use, in this
paper we use the term LLMs to be more specific about what we are focused
on. LLMs have been quickly adopted in education by both instructors and
students. This adoption brings both concerns and opportunities.

The approach of our study is similar to Kross and Guo (2019) who
conducted interviews with 20 data scientists focused on teaching. They
were interested in understanding what is taught, how they teach, and
some of the challenges. Their context was data science more broadly,
whereas our focus is the impact of LLMs.

In terms of LLMs and data science education, Ellis and Slade (2023) and
Tu et al. (2024) both detail how data science educators should, and
should not, use LLMs in data science education. Some of their
recommendations for educators include that educators should use LLMs as
a curriculum and assessment design assistant (Ellis and Slade 2023; Tu
et al. 2024). They also recommend showing how LLMs make mistakes on data
science tasks as an exercise for students to practice critical thinking
(Ellis and Slade 2023; Tu et al. 2024). Ellis and Slade (2023)
recommends that educators should teach and guide students about how to
use LLMs to debug their code. Both Ellis and Slade (2023) and Tu et al.
(2024) argue that educators should change assessment. This may mean
focusing on tasks on which they say LLMs do not do well, such as
interpreting statistical output that is provided as an image or complex
multi-stage tasks. But it also means more oral presentations, and basing
grading and feedback on the process of completing exercises and
projects, not just final outputs. Tu et al. (2024) recommended that
educators should adopt clear LLM use policies. In particular they
suggested educators require that LLM usage be cited and that educators
should provide examples of how to do this. Finally Tu et al. (2024)
recommended that educators should increase the amount of ethics content
in their programs.

Ellis and Slade (2023) and Tu et al. (2024) also had several cautions in
terms of the use of LLMs in data science education. Both were especially
concerned about vigilance with regard to academic integrity. Their main
concern was about how to assess student knowledge. They recommended
testing assessment questions on LLMs before they were provided to
students. This was to see how well LLMs could do on them. They also
recommended blocking access to particular websites to ensure LLMs could
not be used.

Barba (2025) describe their experience of using GenAI in an engineering
course focused on computation. They found that students used AI as a
short-cut rather than as a tool, leading to an ``illusion of
competence''. They recommend the use of active learning and reflective
practices and not just the redesign of assessment.

Tu et al. (2024) felt that LLMs should only be used at the intermediate
and advanced stages of data science education, so as to not interfere
with students learning fundamental skills and concepts. They were also
concerned that introductory-level students may not be able to assess LLM
accuracy and relevance. This is a similar concern to that of Prather et
al. (2024) who find that students who lack metacognitive awareness
(which is the understanding of one's understanding) and use LLMs are put
at a larger disadvantage than before compared to those who have
awareness and use LLMs.

\section{Study design}\label{sec-design}

To study the view of faculty on LLMs we used the qualitative approach of
semi-structured interviews. This research was approved by the University
of Toronto's Research Ethics Board (Protocol: \#00048749).

The approach that we used was:

\begin{enumerate}
\def\labelenumi{\arabic{enumi}.}
\tightlist
\item
  Conduct interviews.
\item
  Transcribe and anonymize the transcripts.
\item
  Create a codebook.
\item
  Code the qualitative aspects.
\item
  Conduct analysis.
\end{enumerate}

We began by emailing 185 people in the statistics and data science
community. We focused on those teaching applied statistics or data
science courses by identifying those involved with Software Carpentry,
the \emph{Journal of Statistics and Data Science Education} (JSDSE)
editorial team, universities with prominent statistics and data science
programs, authors of recently published pedagogy papers or talks in
venues such as \emph{Harvard Data Science Review}, \emph{JSDSE}, recent
or upcoming Joint Statistical Meetings (JSM), prominent blogs, and arXiv
preprints. An example email is available in Appendix~\ref{sec-email}. We
sent these emails during the month-long period from 9 June 2025 to 9
July 2025 with the vast majority sent in the first week.

In total, we conducted 42 30-minute semi-structured interviews. These
happened remotely over Zoom. Respondents were from 33 unique
institutions, across 10 countries. All interviews were conducted by only
one author.

Respondents were first asked about their name, academic background,
years of teaching experience, and some aspects about the courses that
they typically teach. Questions then focused on their general attitude
to LLMs, the models they used, and the extent to which they were using
LLMs in their classes. We were then interested in potential impacts of
LLMs, such as changed grade distributions, student motivation, and the
broader goals of education. After that we asked questions about whether
their assessment had changed or their expected citation of LLM usage. We
were also interested in whether instructors had seen instances of
hallucinations and whether they believed they could tell when LLMs were
being used. Finally, we asked for their recommendations for others that
we should contact. The interview guide is available in
Appendix~\ref{sec-interview}. Not all respondents were asked every
question, as it depended on their answers to earlier questions.

Interviews were recorded, then transcribed using Zoom's
auto-transcription service. One author anonymized the transcripts by
removing details such as names and specifics of education and
appointment, as well as courses taught. Details of the anonymization
approach are provided in Appendix~\ref{sec-anonymization}. During that
process the transcripts were also cleaned up to create appropriate
paragraphing and remove filled pauses. A different author then checked
the transcripts. They were then sent to the respondent who had a chance
to suggest changes to support anonymization, or to withdraw from the
study. No respondent withdrew. Our anonymized transcripts are available
in the repository that supports this paper.

(Two initial pilot interviews were conducted by two authors together
while developing the interview guide, and one additional interview was
conducted but the respondent was not an instructor of a statistics or
data science related course. None of those interviews are counted in our
statistics.)

Some aspects of the analysis were amenable to direct quantification.
Other aspects required qualitative analysis. To do this two authors read
through all the interview transcripts. One author was then responsible
for creating a codebook. To do this that author first read all the
transcripts and noted key themes, and phrases. Then following Tai et al.
(2024) and Liu et al. (2025) that author used OpenAI's O3 model (as at 1
July 2025, o3-2025-04-16) to develop an initial codebook based on ten
randomly selected anonymized transcripts. The author then extensively
refined and updated the codebook by removing, adding or modifying almost
all codes and descriptions. A second author, not involved in the
creation of the codebook, but who had conducted the interviews and read
all the transcripts, then also went through and similarly updated the
codebook. Finally, a third author reviewed the codebook and made further
refinements. Where there were differences the authors discussed and
agreed on an appropriate coding. The final codebook is available in
Appendix~\ref{sec-codebook}. The authors then applied this codebook to
each interview, reviewed the results, and discussed which were the most
important themes. To interact with the OpenAI API (i.e.~when creating
the initial codebook and conducting the initial coding) we used the
programming language Python (Python Software Foundation 2025) and the
pandas (The pandas development team 2020; McKinney 2010) and openai
(OpenAI 2025) Python packages. For each question, the possible coded
outcome for each response are detailed in
Table~\ref{tbl-interview-questions}.

\begin{table}

\caption{\label{tbl-interview-questions}Interview questions and possible
answer options after coding}

\centering{

\begingroup\fontsize{8}{10}\selectfont

\begin{tabular}[t]{>{\raggedright\arraybackslash}p{8cm}>{\raggedright\arraybackslash}p{8cm}}
\toprule
Question & Possible answers after coding\\
\midrule
\cellcolor{gray!10}{Where are you based?} & \cellcolor{gray!10}{Canada, US, Other}\\
Will you please state your academic background? & Biostatistics, Computer Science, Statistics, Other\\
\cellcolor{gray!10}{How long have you been teaching?} & \cellcolor{gray!10}{1–5 years, 6–10 years, 11–20 years, 20+ years}\\
What do you normally teach? & Computer Science, Data Science, Statistics, Other\\
\cellcolor{gray!10}{What is the average size of your classes?} & \cellcolor{gray!10}{1–10 students, 11–20 students, 21–50 students, 51–100 students, 101–200 students, 201–500 students, 500+ students}\\
\addlinespace
What is the typical level? & Undergraduate, Graduate\\
\cellcolor{gray!10}{What are your general views on LLMs?} & \cellcolor{gray!10}{Ethical concerns, LLM capability concerns, Capacity concerns, Learning integrity concerns, Pragmatic, Specific uses, Excitement, Other}\\
What are your views on using LLMs to teach students? & Conflicted, Student engagement, Pragmatic, Personalized tutor, Calculators, Opposed, Unsure, Concerned, Foundations, Other\\
\cellcolor{gray!10}{Have you used LLMs to teach students, if so, how have you used LLMs to teach students?} & \cellcolor{gray!10}{None, Planned, Prep only, Demo use, Guided use, Extensive, Other}\\
What LLM interface do you use to teach students (i.e. ChatGPT, Claude, Gemini, etc.)? & OpenAI, Anthropic, Google, GitHub, Other\\
\addlinespace
\cellcolor{gray!10}{Have you noticed a change in grade distributions pre-AI to now?} & \cellcolor{gray!10}{Inflated, Stable, Bimodal, Compressed, Confounded, Office hours, Other}\\
Do you think there's a difference in the goal of education now vs pre-AI? & Unchanged, Conceptual change, AI-literacy, Assessment redesign, Uncertain, Rethinking, Credentialling, Other\\
\cellcolor{gray!10}{Do you find there's a difference in students' attitudes and motivation towards learning now?} & \cellcolor{gray!10}{Same, Speed, Bimodal, Confounded, Motivation concerns, Office hours, Other}\\
What are the benefits of using LLMs to teach students? & Skills, Engagement, Personalized tutor, Efficiency, Scaling, Other\\
\cellcolor{gray!10}{What concerns do you have of using LLMs to teach students?} & \cellcolor{gray!10}{De-skilling, Illusory mastery, Black-boxes, Shortcuts, Equity, Moving target, Other}\\
\addlinespace
In the past few years, how have you changed your assessment given the ubiquity of LLMs? & In-class return, Reduced take-home weight, Process proof, AI-integrated tasks, Authenticity reduction, No change, Other\\
\cellcolor{gray!10}{In the coming few years, how do you think you will change your assessment?} & \cellcolor{gray!10}{In-class return, Reduced take-home weight, Process proof, AI-integrated tasks, Unsure, No change, Other}\\
Do you ask your students to cite LLMs in their work? & No, Guided, Honor, Yes, Other\\
\cellcolor{gray!10}{Have you seen instances of hallucinations or other made-up aspects in submitted work?} & \cellcolor{gray!10}{Obvious fabrication, Stylistic tells, Self-corrected, Undetected or unsure, No, Other}\\
Putting hallucinations to one side, do you think you can tell when some code or writing has been produced by LLMs? & Obvious fabrication, Pattern spotting, Identical submissions, Asked for explanation, Undetected or unsure, Pointless, No, Other\\
\addlinespace
\cellcolor{gray!10}{How do you think academia will change to integrate LLMs into education?} & \cellcolor{gray!10}{Guided use increase, Instructor to coach change, Assessment changes, Inconsistent policies, Personalized tutor, Decreased degree value, Messy, Concerned, Slowly, Pragmatic, Other}\\
\bottomrule
\end{tabular}
\endgroup{}

}

\end{table}%

Some responses were coded uniquely. For instance, where a respondent was
based, and their amount of experience. But others could have multiple
codes. For instance, some instructors taught classes of different sizes,
or had experience with multiple fields. More importantly, responses to
questions like their opinions about LLMs could have multiple codes. One
response could be coded to multiple themes, so our bar graphs do not
need to add to 100 per cent. Finally, some questions were not asked of
some respondents where their previous responses made it clear it was not
relevant. For instance, if they said they did not use LLMs at all, then
they were not asked which LLM they used.

We aggregated the codes to summarize the responses. To protect
anonymity, for questions about geographic location, academic background
and teaching experience, in cases where codes had less than 3
observations the code was changed to ``Other'' and the identifying
information was redacted from the transcript to the best of our ability.
For all other questions, codes were changed to ``Other'' only if there
was just a single observation for a code. In these cases transcripts
were not redacted as the responses to these questions were viewed as
very low risk for breaking anonymity. To calculate statistical summaries
and create data visualizations, we primarily used the statistical
programming language \texttt{R} (R Core Team 2025), as well as the
\texttt{tidyverse} suite of packages (Wickham et al. 2019).

\section{Results}\label{sec-results}

\begin{figure}

\centering{

\pandocbounded{\includegraphics[keepaspectratio]{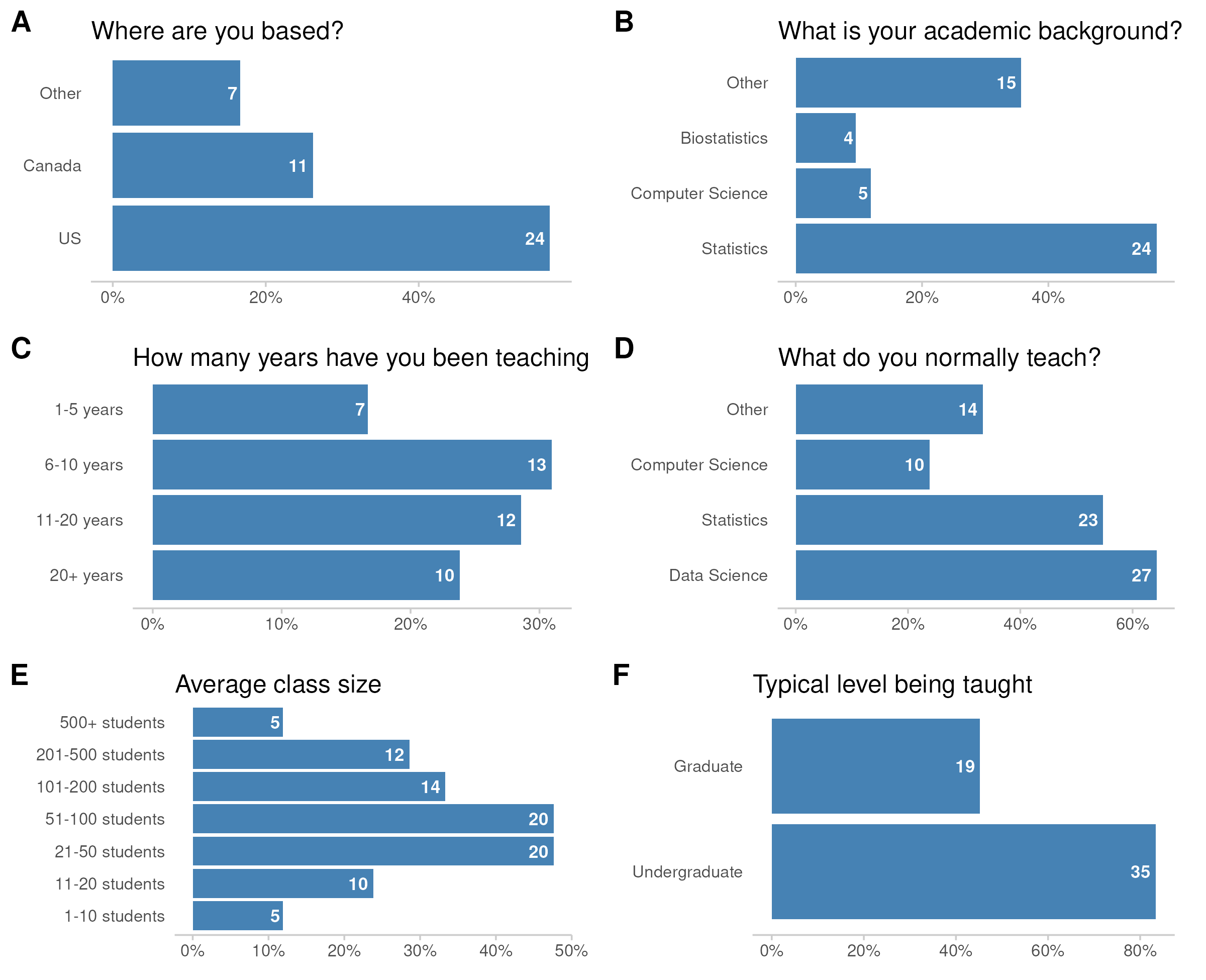}}

}

\caption{\label{fig-general-first}Geographic location, education
background, and teaching experience of the interview respondents. The
percentage and count of each coded theme (bolded text above the bar)
across the interview responses is represented by the blue bar and the
white number, respectively.}

\end{figure}

We first asked our 42 interview respondents questions about their
geographic location, academic background and their teaching experience.
Regarding geographic location, 57 per cent of our respondents were based
in the US, 26 per cent in Canada, and the remainder in other countries
(Figure~\ref{fig-general-first}A). Academic background was typically the
respondent's PhD discipline, or alternatively the discipline of their
highest degree or their postdoctoral studies. A majority had a
background in statistics, with other common disciplines including
computer science, or biostatistics (Figure~\ref{fig-general-first}B).
Respondents typically had considerable experience teaching, with the
most common response being that they had 6-10 years of experience,
followed by 11-20 years, and then 20+ years; only a small number of
respondents had 5 or fewer years of experience
(Figure~\ref{fig-general-first}C). Most respondents taught data science
and statistics --- 64 per cent and 55 per cent, respectively
(Figure~\ref{fig-general-first}D). A smaller group taught
analytics-focused courses in science, or computer science, such as
database design, that are relevant for data scientists. Average class
sizes typically ranged from 21 to 100 students, although a considerable
number reported teaching larger classes of 101 to 500 students
(Figure~\ref{fig-general-first}E). Finally, 83 per cent of respondents
had experience teaching at the undergraduate level, and 45 per cent had
experience teaching at the graduate and/or postgraduate level, including
both for-credit university courses and non-credit workshops. Given that
respondents could realistically report multiple disciplines for their
academic background, teaching topics, average class sizes and
instructional levels, we allowed multiple responses per respondent for
these questions.

Next we asked our respondents 15 questions about their general attitude
to LLMs, as well as their usage and opinions about them within the
context of teaching. For all of the answers to these questions, we
allowed multiple responses to be coded per respondent.

\begin{figure}

\centering{

\pandocbounded{\includegraphics[keepaspectratio]{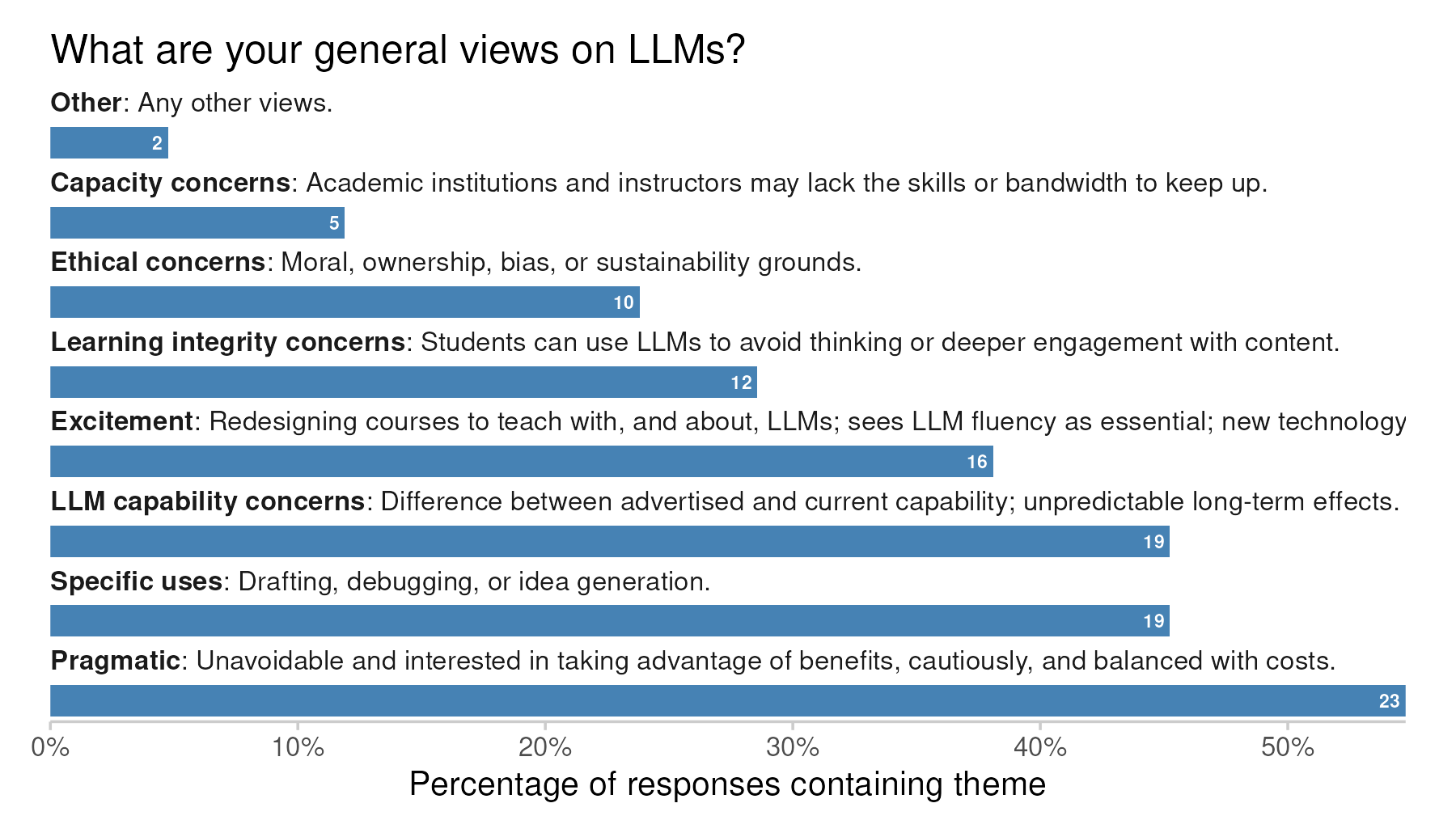}}

}

\caption{\label{fig-general-views}Most frequent themes in interview
responses to the question ``What are your general views on LLMs?'' The
percentage and count of each coded theme (bolded text above the bar)
across the interview responses is represented by the blue bar and the
white number, respectively. Each interview response could contain more
than one theme. A concise description of the coded theme (unbolded text)
follows each coded theme (bolded text).}

\end{figure}

When respondents were asked ``What are your general views on LLMs?'',
seven clear themes emerged (pragmatic, specific uses, LLM capability
concerns, excitement, learning integrity concerns, ethical concerns, and
capacity concerns) with over half of the themes falling under the larger
theme of having concerns about LLMs and their usage
(Figure~\ref{fig-general-views}). Many responses were thoughtful and
complex, and as a consequence were coded into more than one of these.
For example, the quote below was coded as ``pragmatic'', ``specific
uses'', ``learning integrity concerns'', and ``capacity concerns''.

\begin{quote}
\emph{My general views on LLMs are there are pros and cons to them. I am
mostly mindful of it in the context of my teaching. I look favorably on
LLMs for efficient use of my own work. But I am skeptical of its usage
amongst my students broadly. I don't know how I feel, or I don't really
like how ubiquitous they have become, so quickly. I think if there was a
more gradual transition to this, maybe that would be better, but I just
feel like the technological revolution has just been happening so fast,
and then LLMs feel like they've been happening even faster. So, it's
just been a hard transition for me as an educator.}
\end{quote}

The most frequent theme from the answers to this question was pragmatic
(55 per cent), which we defined as thinking that LLMs are unavoidable
and interested in taking advantage of benefits, cautiously, and balanced
with costs. This quote is representative of the responses we coded under
the theme ``pragmatic'', illustrating the practical and realistic
thinking that was common among these responses.

\begin{quote}
\emph{I was very reluctant to do that to begin with, but I'm reaching
the conclusion that this is a technology that isn't going to go
anywhere, and it's better to figure out for myself what they're really
capable of.}
\end{quote}

The next most frequent theme was specific uses (45 per cent), which we
defined as using LLMs tactically for drafting, debugging, or idea
generation. As the following quote demonstrates, responses categorized
under ``specific uses'' frequently conveyed that participants did not
consider LLMs appropriate for every task.

\begin{quote}
\emph{I think they are great as assistance. I think they're great at
helping people organize ideas or have a conversation partner. They're
great for brainstorming. I am not as keen on them for generating content
and different purposes.}
\end{quote}

The third most frequent theme was concerns about LLM capability (45 per
cent), which we defined as concerns about the difference between
advertised and current capability as well as worry about unpredictable
long-term effects. There was also some excitement about LLMs, as it was
the fourth most frequent theme (38 per cent). While we observed there
were many different ways that respondents were concerned, those that
were excited tended to be excited about similar aspects. For instance,
respondents were excited about productivity improvement especially with
regard to coding, they were also excited about the potential for LLMs as
personalized tutors, and they were excited about the emergence of a new
technology.

\begin{figure}

\centering{

\pandocbounded{\includegraphics[keepaspectratio]{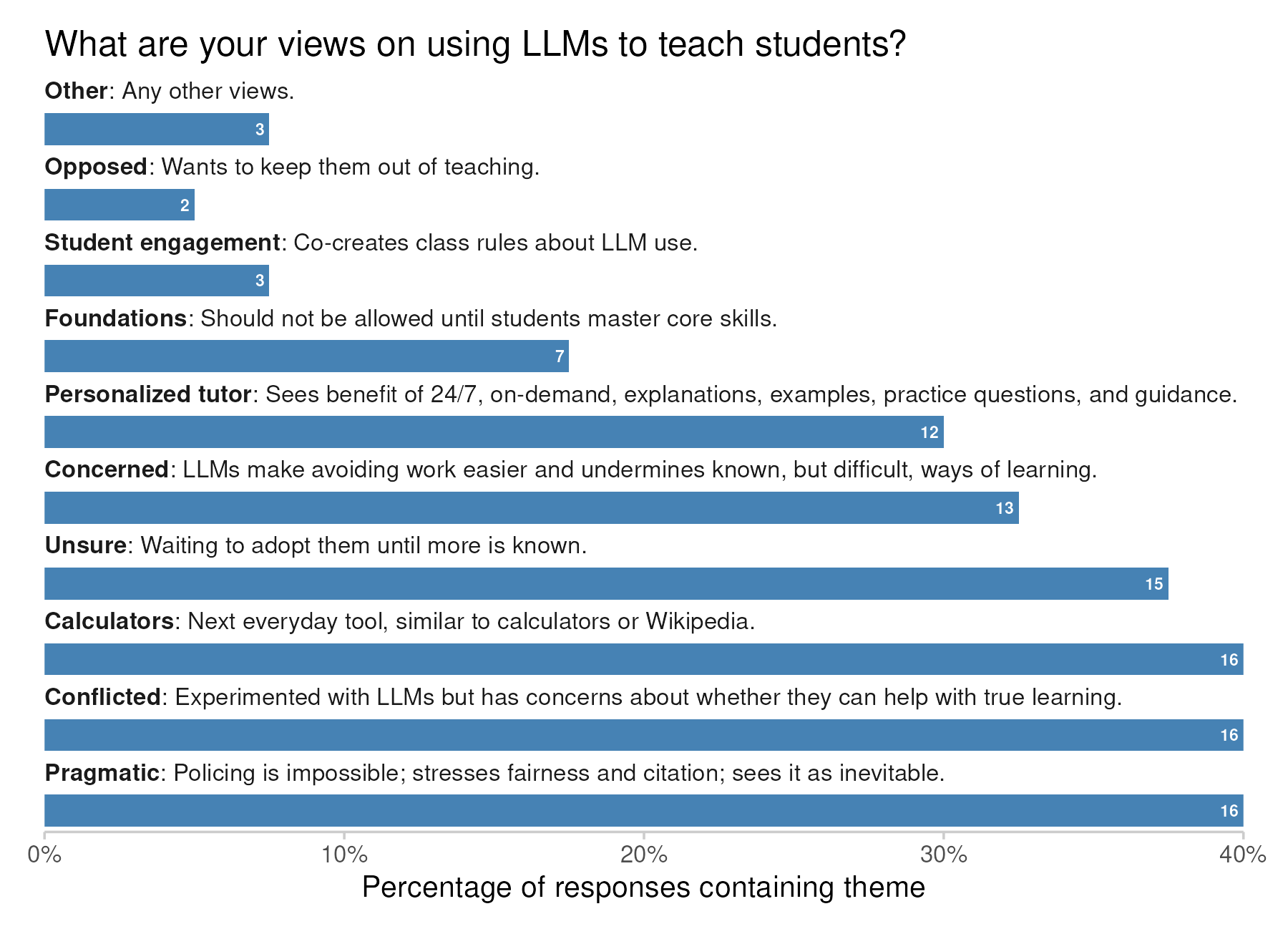}}

}

\caption{\label{fig-teaching-views}Most frequent themes in interview
responses to the question ``What are your views on using LLMs to teach
students?'' The percentage and count of each coded theme (bolded text
above the bar) across the interview responses is represented by the blue
bar and the white number, respectively. Each interview response could
contain more than one theme. A concise description of the coded theme
(unbolded text) follows each coded theme (bolded text).}

\end{figure}

For the next question we asked, ``What are your views on using LLMs to
teach students?'', nine themes were identified --- pragmatic,
conflicted, calculators, unsure, concerned, personalized tutor,
foundations, student engagement, and opposed --- with over half of the
themes having neutral or negative sentiments
(Figure~\ref{fig-teaching-views}). Typical phrases associated with those
less positive responses included the breakdown between traditional
learning and the associated activities. For instance:

\begin{quote}
\emph{I think the crux of the problem\ldots{} is that for essentially
all of human history until now, a big part of helping people learn is,
read this thing, think about it, and then we can talk about it, and then
write something about it to demonstrate that you understood what was in
there. And that basic set of activities has governed virtually all of
what we think of as human learning.}
\end{quote}

Three themes tied as being the most frequent in the responses to this
question (each 40 per cent): ``pragmatic'' (allows LLMs because policing
is impossible; stresses fairness and citation; sees it as inevitable and
that trying to stop it is just swimming against the tide),
``conflicted'' (has experimented with LLMs but has concerns about
whether they can help with true learning) and ``calculators'' (thinks of
LLMs as the next everyday tool, similar to calculators or Wikipedia).
The theme ``unsure'' (is waiting to adopt them until more is known) was
also very frequent, occurring in 38 per cent of responses. The quote
below highlights that such uncertainty may stem from a lack of evidence
about the utility of using LLMs in teaching and a desire for more
research in this area.

\begin{quote}
\emph{But I've been wondering, is there research that tries to evaluate
outcomes and effects {[}of using LLMs{]} on students? There is some, but
I think not nearly enough yet, and it's kind of scattered.}
\end{quote}

\begin{figure}

\centering{

\pandocbounded{\includegraphics[keepaspectratio]{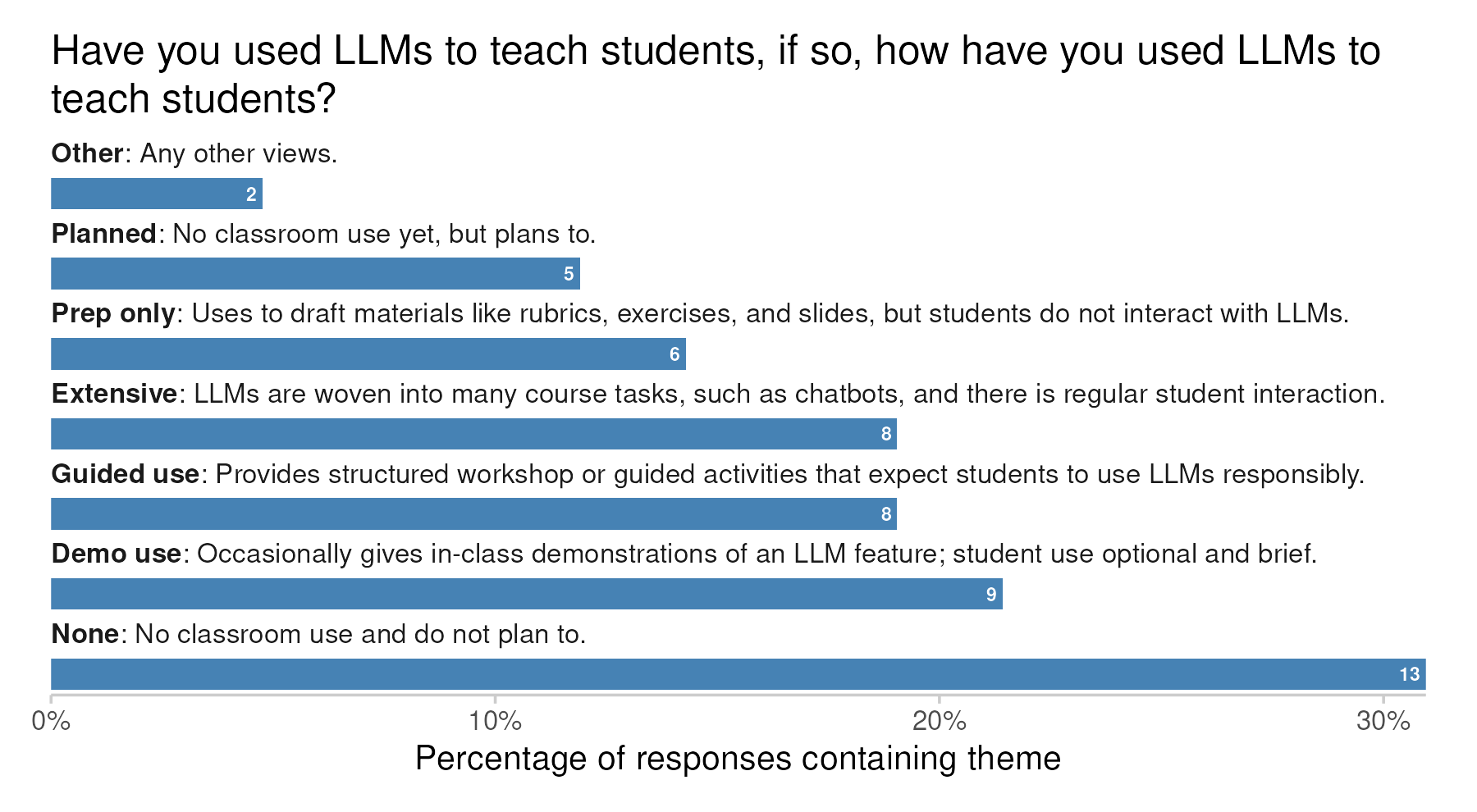}}

}

\caption{\label{fig-how-used}Most frequent themes in interview responses
to the question ``Have you used LLMs to teach students, if so, how have
you used LLMs to teach students?'' The percentage and count of each
coded theme (bolded text above the bar) across the interview responses
is represented by the blue bar and the white number, respectively. Each
interview response could contain more than one theme. A concise
description of the coded theme (unbolded text) follows each coded theme
(bolded text).}

\end{figure}

\begin{figure}

\centering{

\pandocbounded{\includegraphics[keepaspectratio]{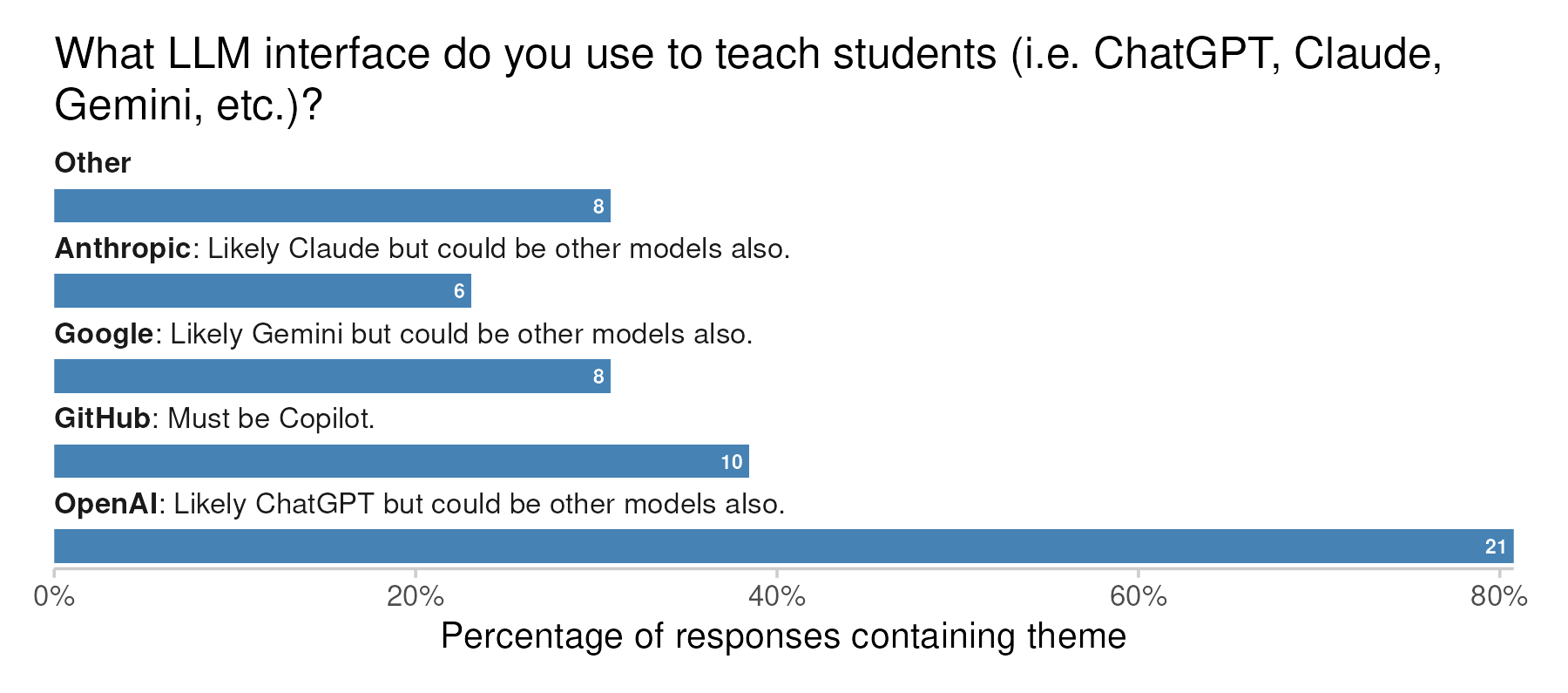}}

}

\caption{\label{fig-which-llm}Most frequent themes in interview
responses to the question ``What LLM interface do you use to teach
students (i.e.~ChatGPT, Claude, Gemini, etc.)?'' The percentage and
count of each coded theme (bolded text above the bar) across the
interview responses is represented by the blue bar and the white number,
respectively. Each interview response could contain more than one theme.
A concise description of the coded theme (unbolded text) follows each
coded theme (bolded text).}

\end{figure}

Six prominent themes were drawn from responses to the question ``Have
you used LLMs to teach students, if so, how have you used LLMs to teach
students?'': none, demonstration use only, guided use, extensive use,
only being used for preparation, and planned but not current use
(Figure~\ref{fig-how-used}). Responses to this question indicated that
many were not using LLMs to teach students (31 per cent). The most
common use was for demonstration purposes (21 per cent), followed by
guided (19 per cent) and extensive use (19 per cent). Fewer used it only
for teaching preparation (14 per cent) or had not yet used it, but
planned to use it in the future (12 per cent). The majority of those who
use LLMs for teaching, use OpenAI's ChatGPT (81 per cent;
Figure~\ref{fig-which-llm}). Microsoft GitHub Copilot was also mentioned
commonly (38 per cent), and often because it was provided by their
institution.

\begin{figure}

\centering{

\pandocbounded{\includegraphics[keepaspectratio]{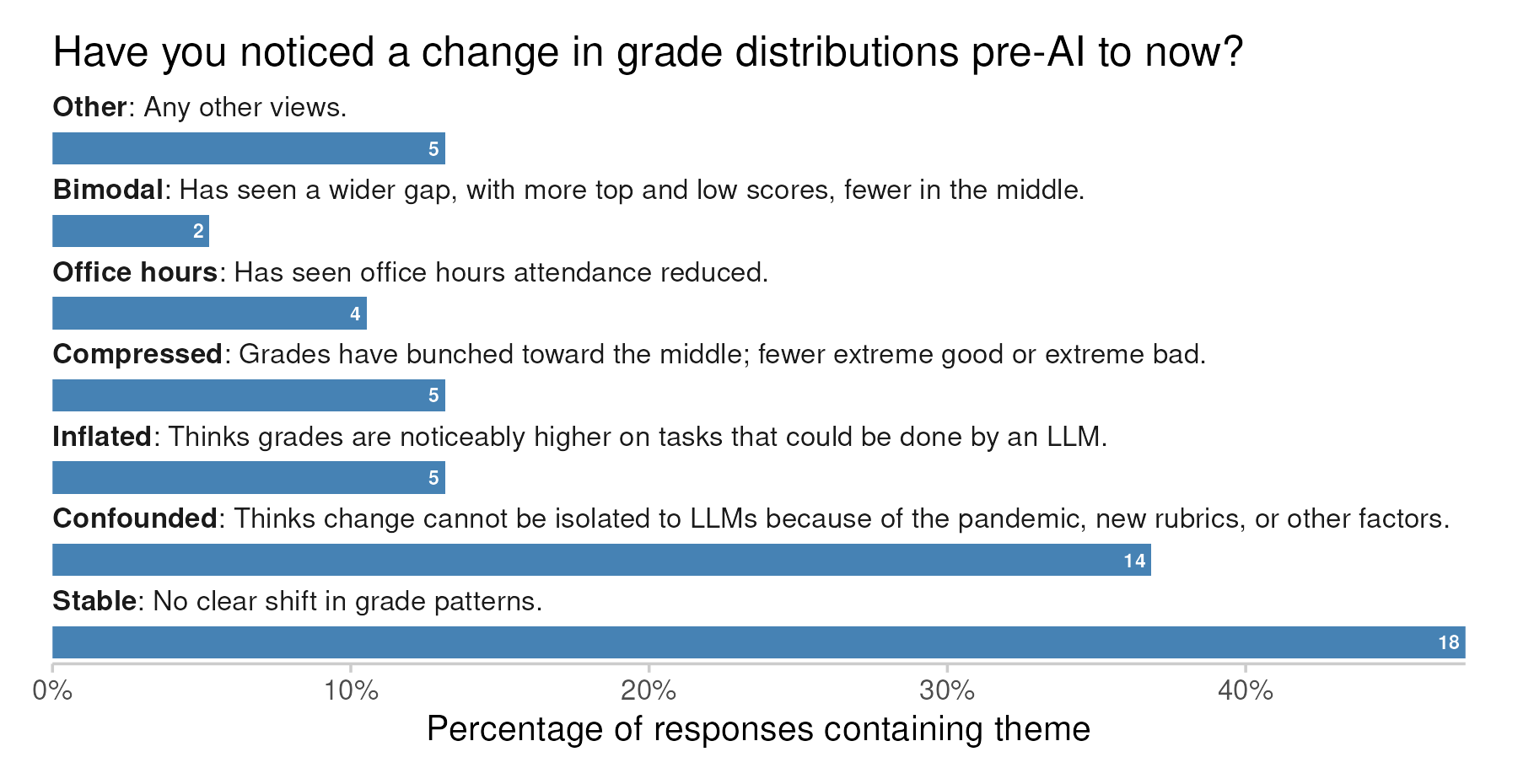}}

}

\caption{\label{fig-grade-change}Most frequent themes in interview
responses to the question ``Have you noticed a change in grade
distributions pre-AI to now?'' The percentage and count of each coded
theme (bolded text above the bar) across the interview responses is
represented by the blue bar and the white number, respectively. Each
interview response could contain more than one theme. A concise
description of the coded theme (unbolded text) follows each coded theme
(bolded text).}

\end{figure}

The question ``Have you noticed a change in grade distributions pre-AI
to now?'' prompted responses that revealed six recurring themes (stable,
confounded, inflated, compressed, office hours, and bimodal)
(Figure~\ref{fig-grade-change}). Although the uniquely coded themes were
evenly split between those clearly indicating a change in grades and
those indicating that it was not clear, or there was not a change, the
most frequent themes in the responses were ``stable'', (47 per cent; no
clear shift in grade patterns since LLMs became widely available) and
``confounded'' (37 per cent; thinks change cannot be isolated to LLMs
because of the pandemic, new rubrics, or other factors). A number of
respondents' responses indicated they thought office hours attendance
had decreased (11 per cent). They tended to attribute this to LLMs, and
believed that it suggested students were able to get responsive help
when they needed it. There was some concern however, that students would
miss out on a human connection. Those who did think that marks had
changed mentioned the difference between results on assessment conducted
under exam conditions and those where LLM usage is possible.

\begin{figure}

\centering{

\pandocbounded{\includegraphics[keepaspectratio]{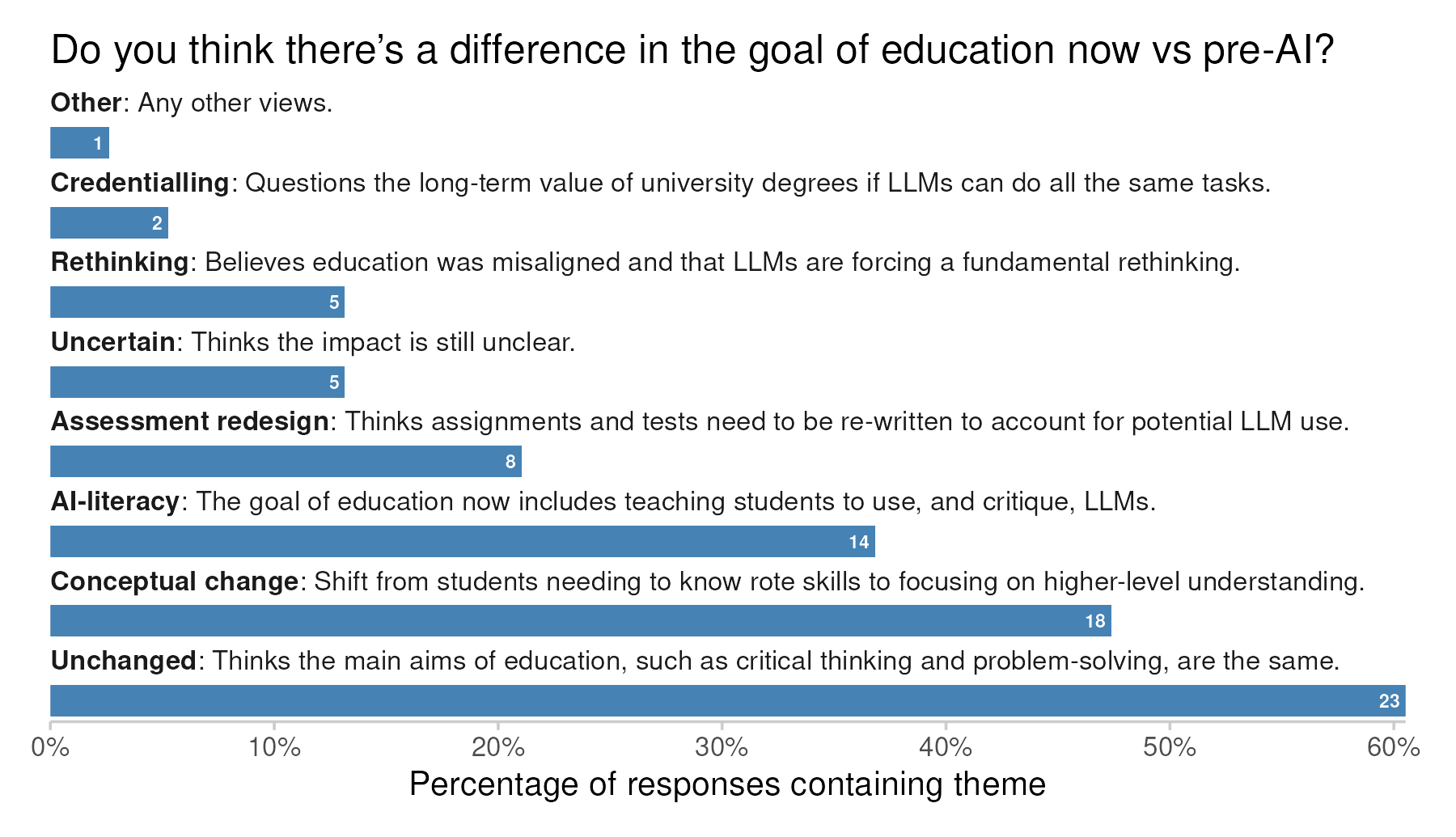}}

}

\caption{\label{fig-changed-goals}Most frequent themes in interview
responses to the question ``Do you think there's a difference in the
goal of education now vs pre-AI?'' The percentage and count of each
coded theme (bolded text above the bar) across the interview responses
is represented by the blue bar and the white number, respectively. Each
interview response could contain more than one theme. A concise
description of the coded theme (unbolded text) follows each coded theme
(bolded text).}

\end{figure}

We were also interested in potential changes in education goals, and
next asked ``Do you think there's a difference in the goal of education
now vs pre-AI?'' In the responses to this question, 7 key themes became
apparent (unchanged, conceptual change, AI-literacy, assessment
redesign, uncertain, rethinking and credentialing). Most respondents (61
per cent), felt that the goal of education was unchanged now compared
with pre-AI (Figure~\ref{fig-changed-goals}). Those who thought that the
goal had changed tended to think that there had been a conceptual change
(47 per cent), or that AI-literacy was now important (37 per cent). Many
(21 per cent) were thinking about redesigning their assessment. This
question also resulted in a considerable number of reflections on the
part of many respondents. For instance:

\begin{quote}
\emph{I love this question because it made me both very happy and very
sad to think about. So, on some level, what is the actual mission and
purpose? Like, why am I a teacher? What am I here for? For these
students? That hasn't changed at all. And I can send you the link to it,
but I also looked up, we have, at the {[}names institution{]}, this
graduate profile, our kind of aspirational statement, of what we think
all students who have left us with a degree should have. And it includes
things like being scholars, they're knowledgeable and curious, excited
by ideas, and conscientious in their efforts to understand the
complexities of their communities and the world. And I love that.}
\end{quote}

\begin{figure}

\centering{

\pandocbounded{\includegraphics[keepaspectratio]{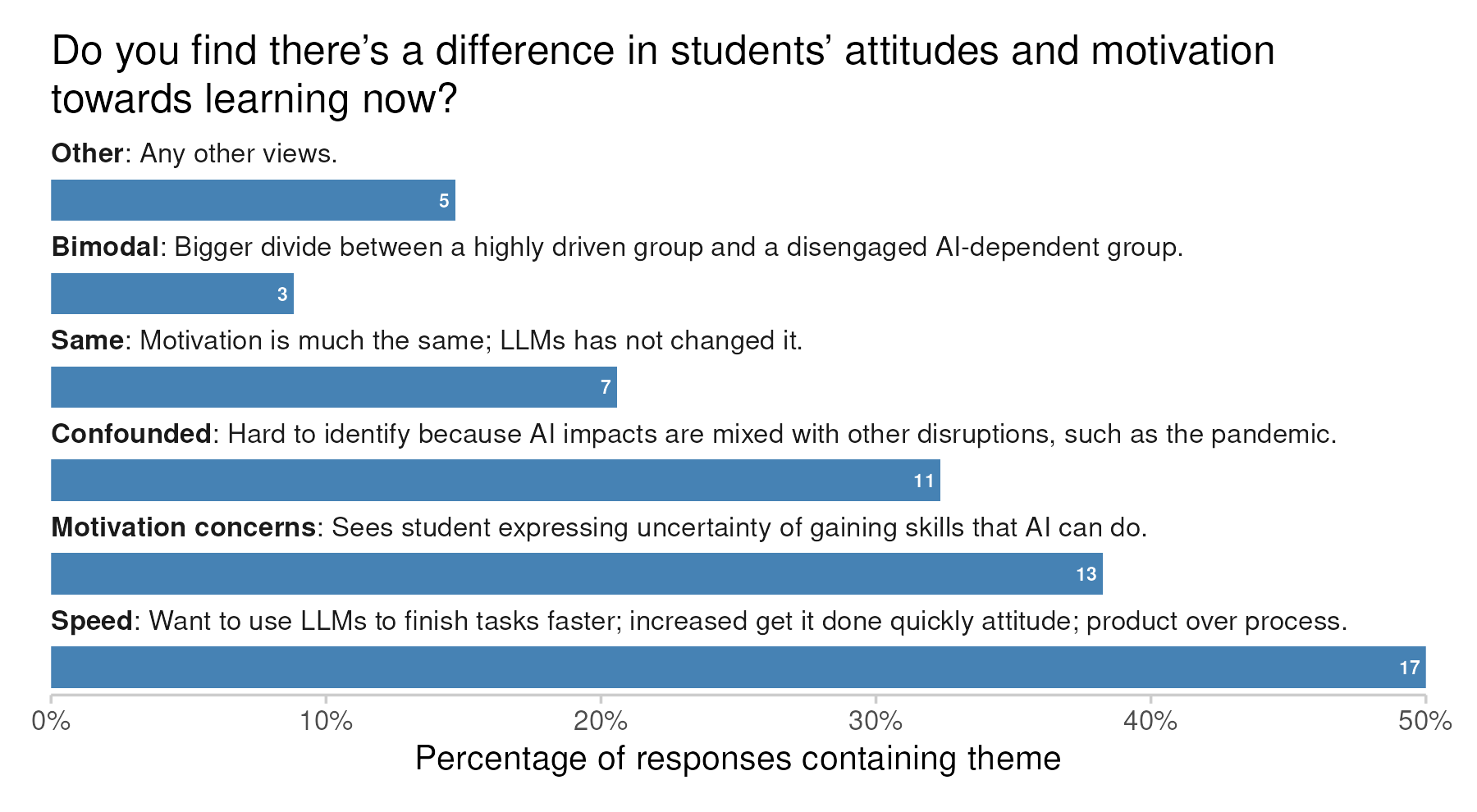}}

}

\caption{\label{fig-changed-attitudes}Most frequent themes in interview
responses to the question ``Do you find there's a difference in
students' attitudes and motivation towards learning now?'' The
percentage and count of each coded theme (bolded text above the bar)
across the interview responses is represented by the blue bar and the
white number, respectively. Each interview response could contain more
than one theme. A concise description of the coded theme (unbolded text)
follows each coded theme (bolded text).}

\end{figure}

The question ``Do you find there's a difference in students' attitudes
and motivation towards learning now?'' sparked responses that mapped
onto six thematic categories --- speed, motivation concerns, confounded,
same, bimodal. Many mentioned a focus on speed (50 per cent) or that
there were concerns about motivation (38 per cent)
(Figure~\ref{fig-changed-attitudes}). That said, a considerable number
(32 per cent) felt that any change due to LLMs would be confounded with
other changes and so be difficult to tease out.

\begin{figure}

\centering{

\pandocbounded{\includegraphics[keepaspectratio]{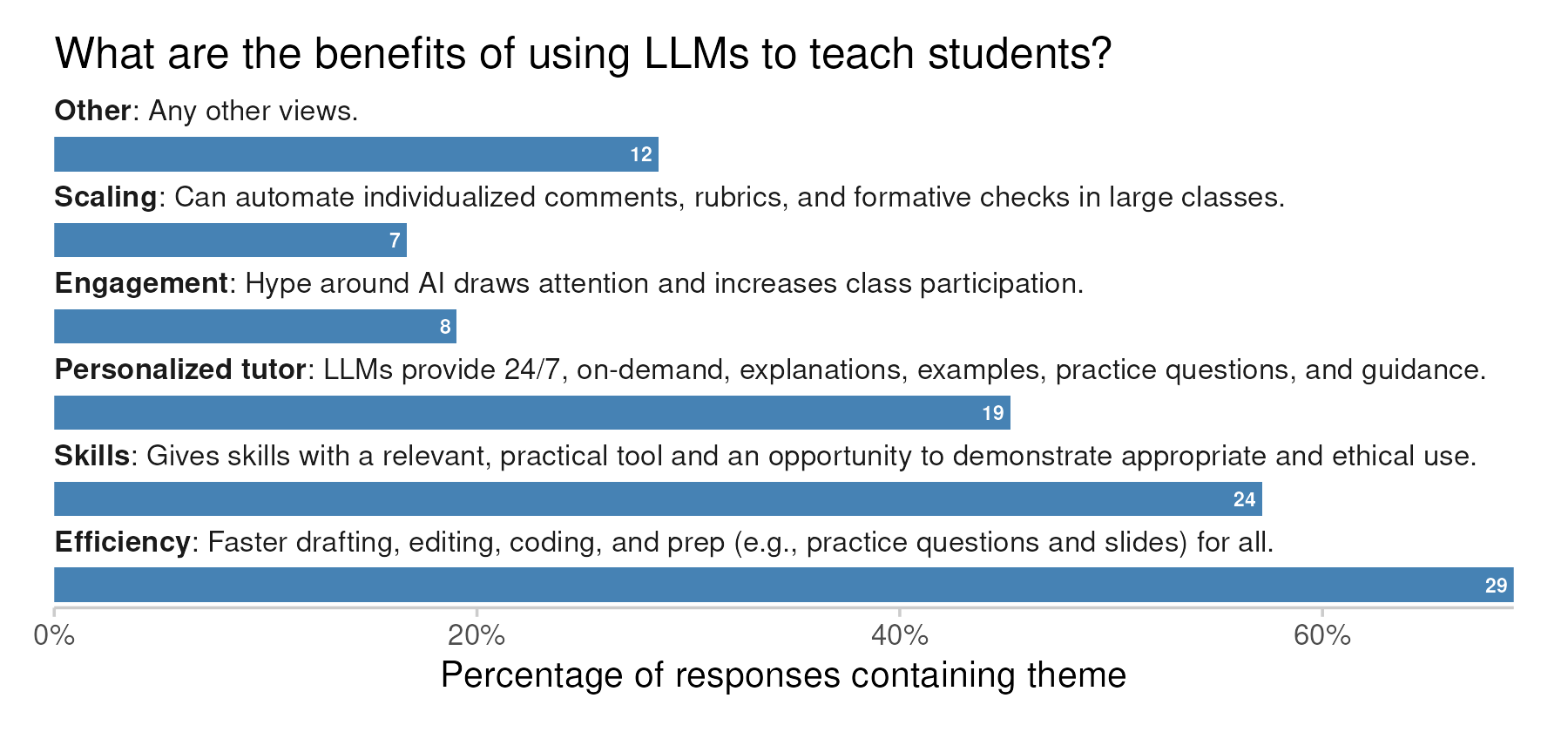}}

}

\caption{\label{fig-benefits}Most frequent themes in interview responses
to the question ``What are the benefits of using LLMs to teach
students?'' The percentage and count of each coded theme (bolded text
above the bar) across the interview responses is represented by the blue
bar and the white number, respectively. Each interview response could
contain more than one theme. A concise description of the coded theme
(unbolded text) follows each coded theme (bolded text).}

\end{figure}

\begin{figure}

\centering{

\pandocbounded{\includegraphics[keepaspectratio]{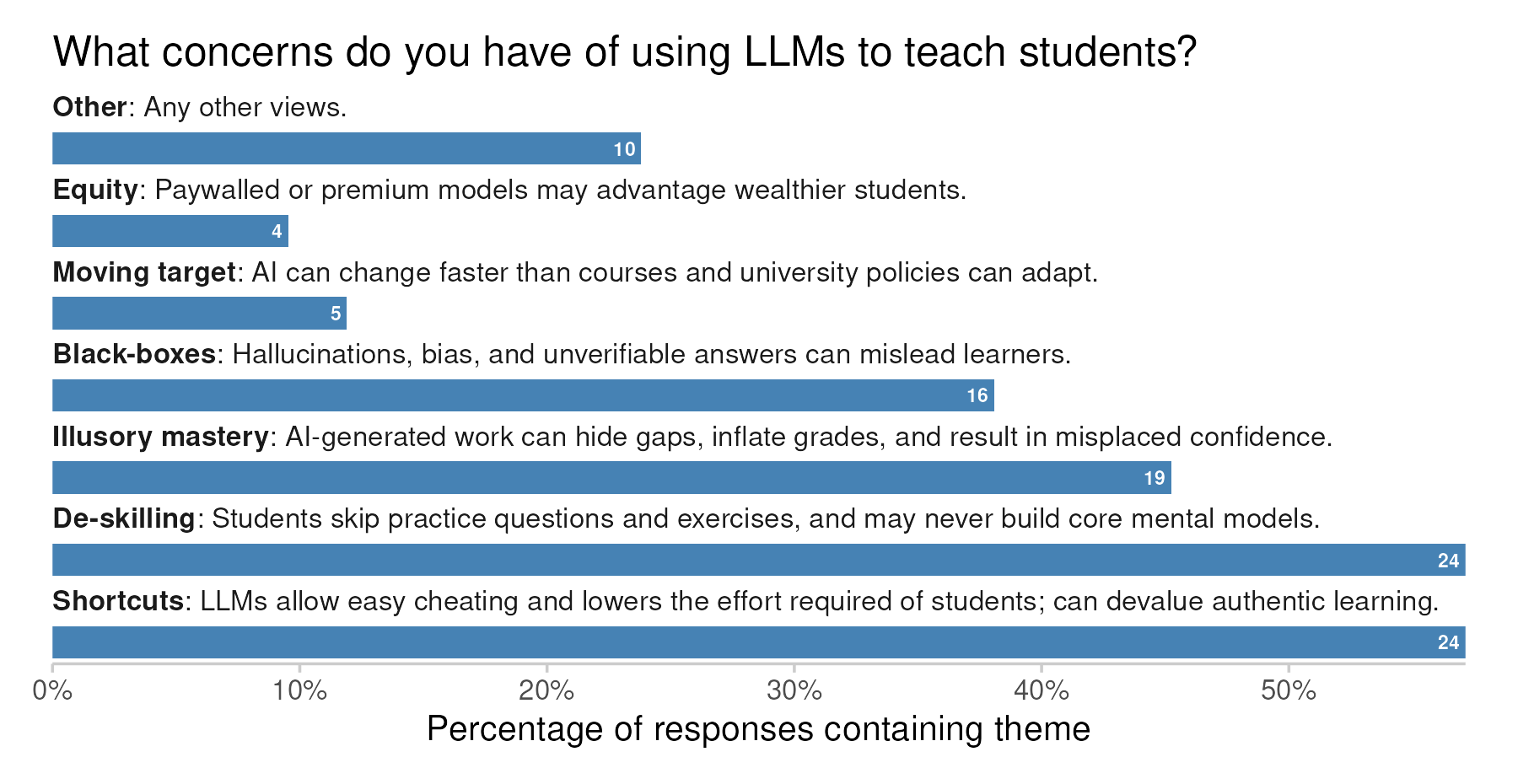}}

}

\caption{\label{fig-concerns}Most frequent themes in interview responses
to the question ``What concerns do you have of using LLMs to teach
students?'' The percentage and count of each coded theme (bolded text
above the bar) across the interview responses is represented by the blue
bar and the white number, respectively. Each interview response could
contain more than one theme. A concise description of the coded theme
(unbolded text) follows each coded theme (bolded text).}

\end{figure}

We were interested in understanding instructors' views on both the
benefits and concerns of using LLMs in data science teaching, and asked
about each explicitly. Responses to ``What are the benefits of using
LLMs to teach students?'' were categorized into 5 major themes:
efficiency, skills, personalized tutor, engagement and scaling.
Respondents felt that there were many benefits of using LLMs to teach
students, including efficiency (69 per cent), the need to develop skills
related to using LLMs (57 per cent) and the potential for LLMs to act as
a personalized tutor (45 per cent) (Figure~\ref{fig-benefits}). From the
responses to ``What concerns do you have of using LLMs to teach
students'', six themes took shape (Figure~\ref{fig-concerns}). These
included the use of LLMs as shortcuts (57 per cent), and the potential
for de-skilling (57 per cent). Another major concern was that students
would incorrectly think that they had mastered something when they had
not (45 per cent), and that LLMs are a black-box (38 per cent).

\begin{figure}

\centering{

\pandocbounded{\includegraphics[keepaspectratio]{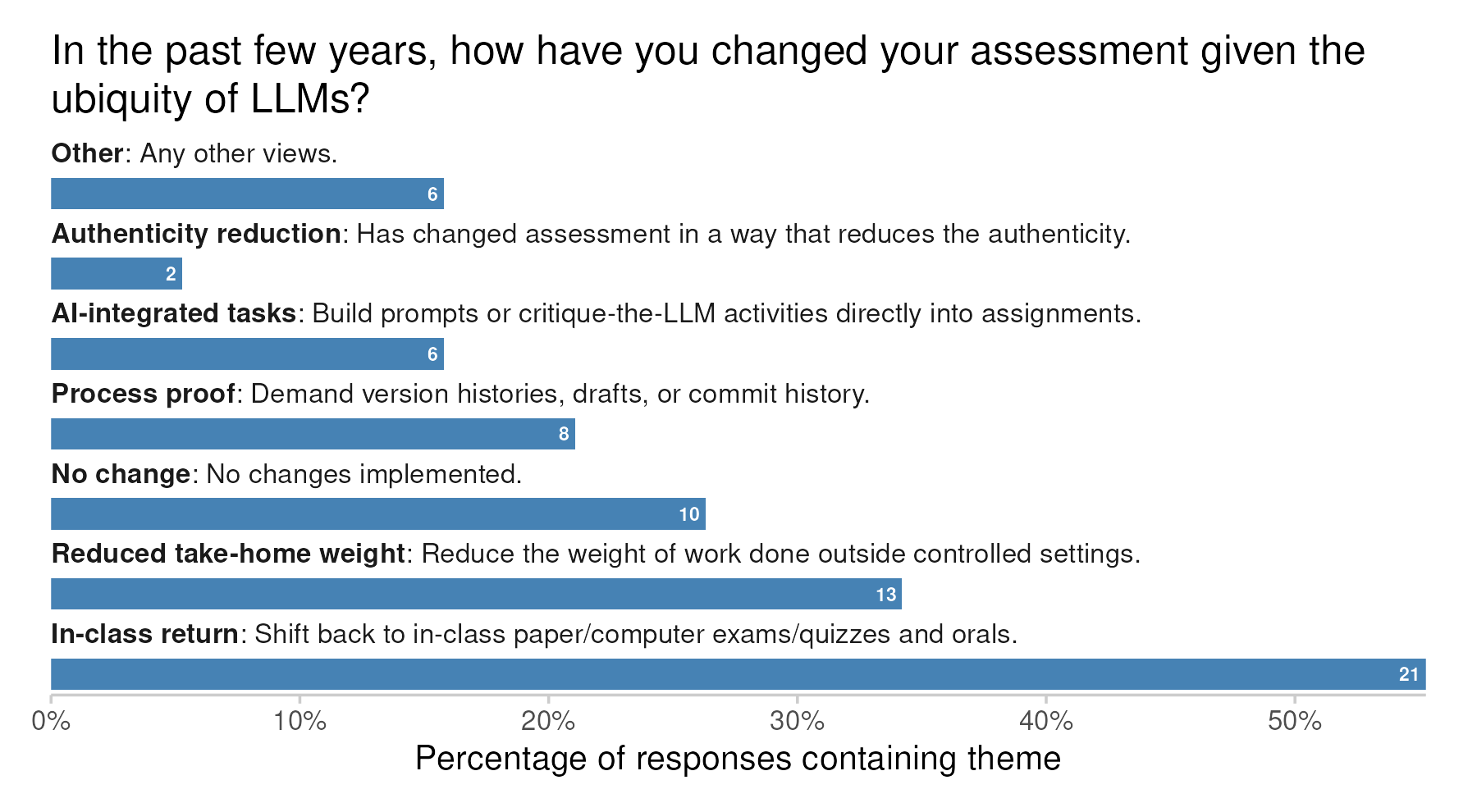}}

}

\caption{\label{fig-has-changed-assessment}Most frequent themes in
interview responses to the question ``In the past few years, how have
you changed your assessment given the ubiquity of LLMs?'' The percentage
and count of each coded theme (bolded text above the bar) across the
interview responses is represented by the blue bar and the white number,
respectively. Each interview response could contain more than one theme.
A concise description of the coded theme (unbolded text) follows each
coded theme (bolded text).}

\end{figure}

We then explored how instructors have adapted, or plan to adapt, their
assessment practices in response to the emergence of LLMs, and asked
about both current and anticipated changes
(Figure~\ref{fig-has-changed-assessment}). When respondents were asked
``In the past few years, how have you changed your assessment given the
ubiquity of LLMs?'' six key themes became apparent: increasing in-class
assessment (55 per cent), reducing the weight of take-home tasks (34 per
cent), making no change (26 per cent), or requiring proof of process (21
per cent).

\begin{figure}

\centering{

\pandocbounded{\includegraphics[keepaspectratio]{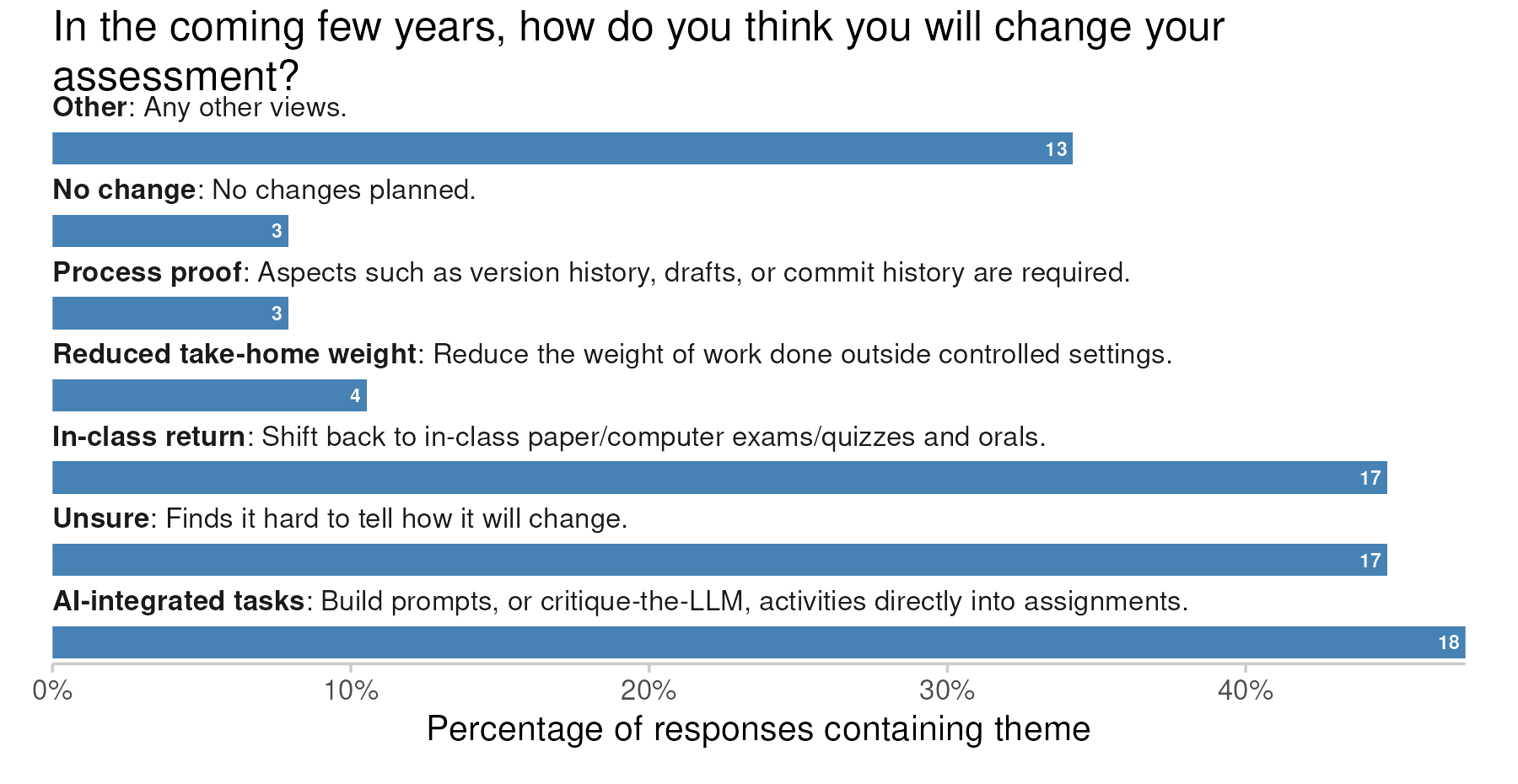}}

}

\caption{\label{fig-will-change-assessment}Most frequent themes in
interview responses to the question ``In the coming few years, how do
you think you will change your assessment?'' The percentage and count of
each coded theme (bolded text above the bar) across the interview
responses is represented by the blue bar and the white number,
respectively. Each interview response could contain more than one theme.
A concise description of the coded theme (unbolded text) follows each
coded theme (bolded text).}

\end{figure}

Similarly, six themes emerged from the responses to ``In the coming few
years, how do you think you will change your assessment?''
(Figure~\ref{fig-will-change-assessment}). Interestingly, there was a
degree of positivity about the future. In particular many (47 per cent)
expected to bring integrated tasks into assessment requirements.
Although there were also many who felt unsure about what they would do
(45 per cent) or expected to return to in-class assessment (45 per
cent). A common approach was to divide assessment into two streams: that
which was conducted in a way that an LLM could not help, such as
proctored exams written on paper, or on computers with internet
controls, and that which could be completed by an LLM, such as take-home
assignments, and to then adjust the expectations and weighting between
the two.

\begin{figure}

\centering{

\pandocbounded{\includegraphics[keepaspectratio]{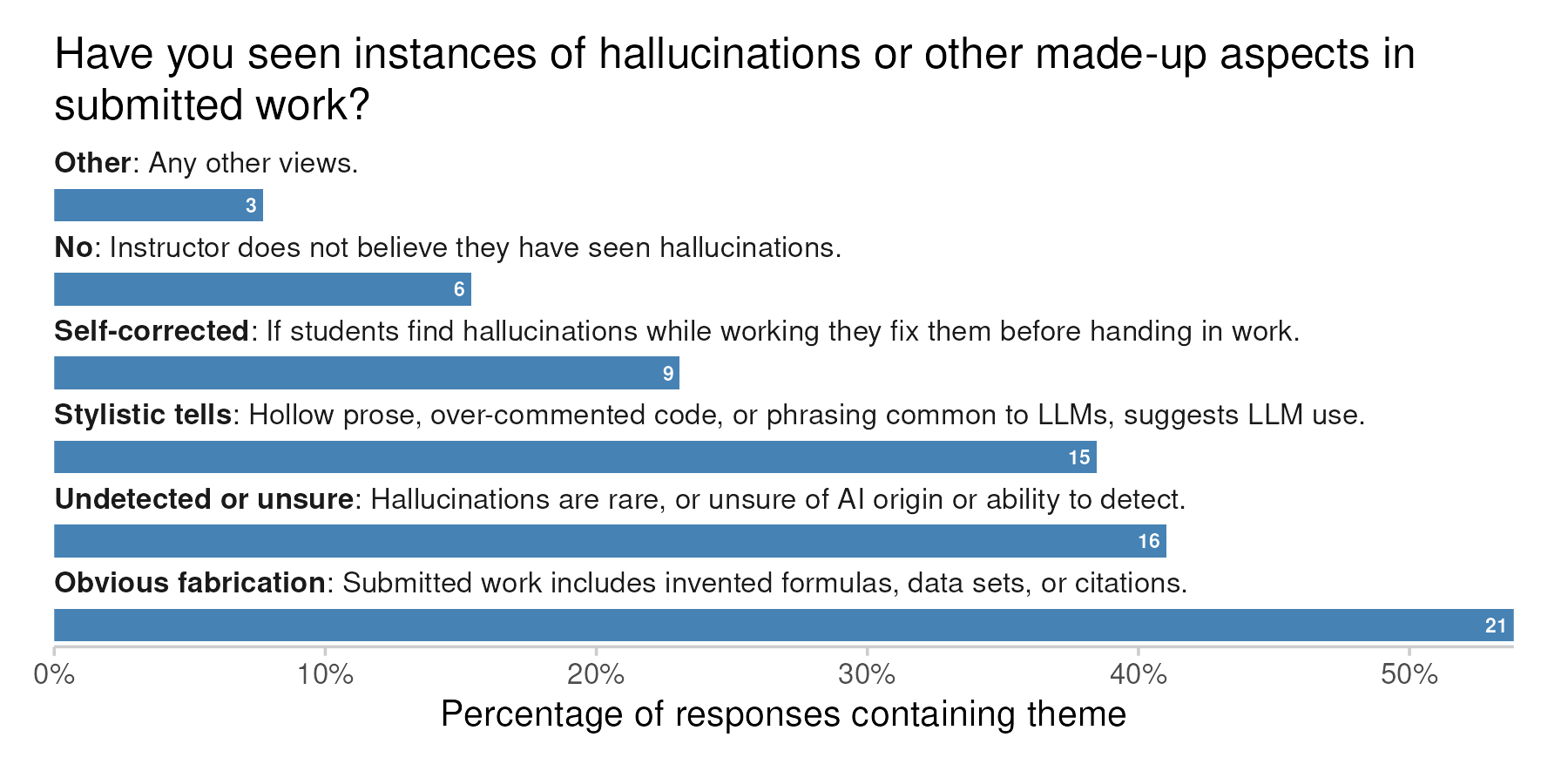}}

}

\caption{\label{fig-seen-hallucinations}Most frequent themes in
interview responses to the question ``Have you seen instances of
hallucinations or other made-up aspects in submitted work?'' The
percentage and count of each coded theme (bolded text above the bar)
across the interview responses is represented by the blue bar and the
white number, respectively. Each interview response could contain more
than one theme. A concise description of the coded theme (unbolded text)
follows each coded theme (bolded text).}

\end{figure}

Given that LLMs can fabricate answers (often called hallucinations
(Maleki, Padmanabhan, and Dutta 2024; Kalai et al. 2025)), we wanted to
know if instructors had observed these in student work. When respondents
were asked, ``Have you seen instances of hallucinations or other made-up
aspects in submitted work'', they expressed a range of experiences that
grouped into five themes: obvious fabrication, undetected or unsure,
stylistic tells, self-corrected, and no. A considerable number (54 per
cent) of respondents have seen obvious fabrication or believed there
were stylistic tells from work done by LLMs (38 per cent) in submitted
work (Figure~\ref{fig-seen-hallucinations}).

\begin{figure}

\centering{

\pandocbounded{\includegraphics[keepaspectratio]{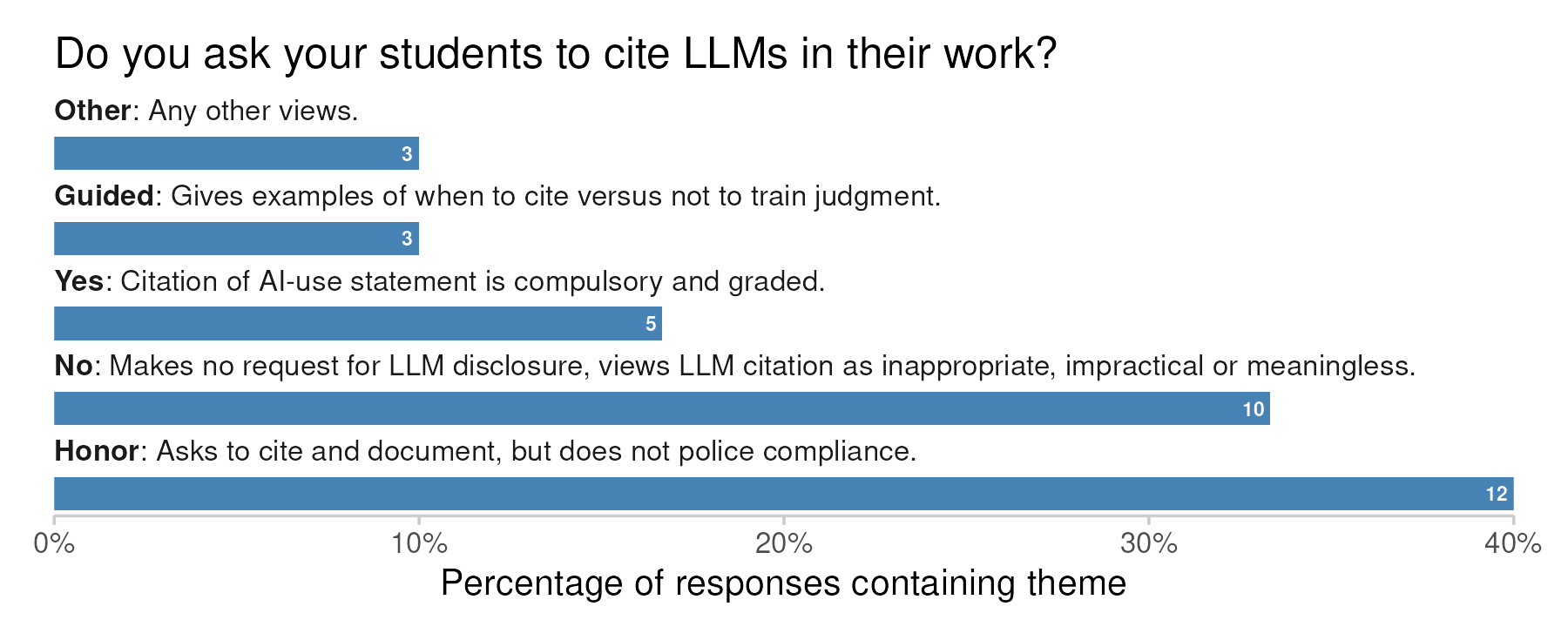}}

}

\caption{\label{fig-cite-llms}Most frequent themes in interview
responses to the question ``Do you ask your students to cite LLMs in
their work?'' The percentage and count of each coded theme (bolded text
above the bar) across the interview responses is represented by the blue
bar and the white number, respectively. Each interview response could
contain more than one theme. A concise description of the coded theme
(unbolded text) follows each coded theme (bolded text).}

\end{figure}

We were interested in how teaching practices are changing in response to
LLMs, particularly in relation to academic integrity. To explore this,
we asked instructors about their policies on citing LLMs and whether
they feel able to identify work generated by these tools. Analysis of
responses to ``Do you ask your students to cite LLMs in their work''
highlighted four central themes (honor, no, yes, guided). Most
respondents either do not ask for citation or have informal citation
expectations of LLM usage, with fewer requiring formal disclosure
(Figure~\ref{fig-cite-llms}). As is often the case with citations, the
citation of LLM usage relies on an honor system (40 per cent). But a
substantial number (33 per cent) said either that they did not, or that
there was no point.

\begin{figure}

\centering{

\pandocbounded{\includegraphics[keepaspectratio]{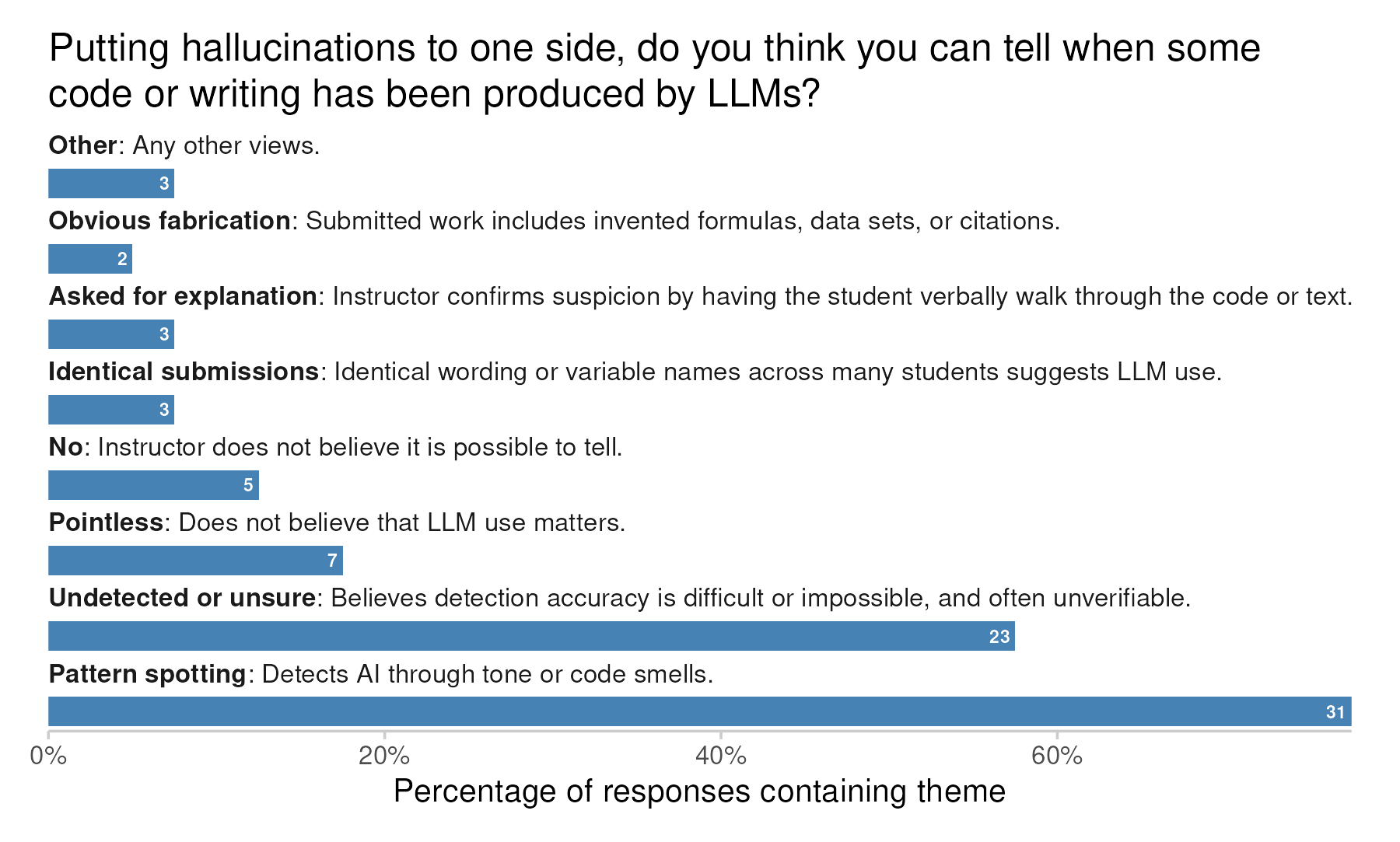}}

}

\caption{\label{fig-can-tell}Most frequent themes in interview responses
to the question ``Putting hallucinations to one side, do you think you
can tell when some code or writing has been produced by LLMs?'' The
percentage and count of each coded theme (bolded text above the bar)
across the interview responses is represented by the blue bar and the
white number, respectively. Each interview response could contain more
than one theme. A concise description of the coded theme (unbolded text)
follows each coded theme (bolded text).}

\end{figure}

When respondents were asked ``Putting hallucinations to one side, do you
think you can tell when some code or writing has been produced by
LLMs?'', seven clear themes were observed (pattern spotting, undetected
or unsure, pointless, no, identical submissions, asked for explanation,
obvious fabrication) (Figure~\ref{fig-can-tell}). Pattern spotting is
the main detection method (78 per cent). But many instructors remain
unsure or unable to reliably tell (57 per cent) and still others (18 per
cent) felt that it was pointless to try. Some respondents described it
being easy to tell when code had been generated by an LLM and difficult
to tell when writing had been generated by an LLM, whereas others felt
the opposite. In terms of writing, many who mentioned this described a
hollowness to the writing. Some felt there was no use being worried
about whether an LLM had been used, and that they instead tried to focus
on other aspects such as whether it was actually good code or writing.
The general feeling about this question was one of pointlessness. For
instance,

\begin{quote}
\emph{I suspect, but I don't know if I'm right or not. Sometimes I
definitely suspect. But I also wonder if how people genuinely write is
influenced in a circular way by what they see out of LLMs, so, part of
me wants to say yes, but also I could be wrong. And I think that anyone
who says yes or no to this question, I feel like we don't know, that's
the point.}
\end{quote}

\begin{figure}

\centering{

\pandocbounded{\includegraphics[keepaspectratio]{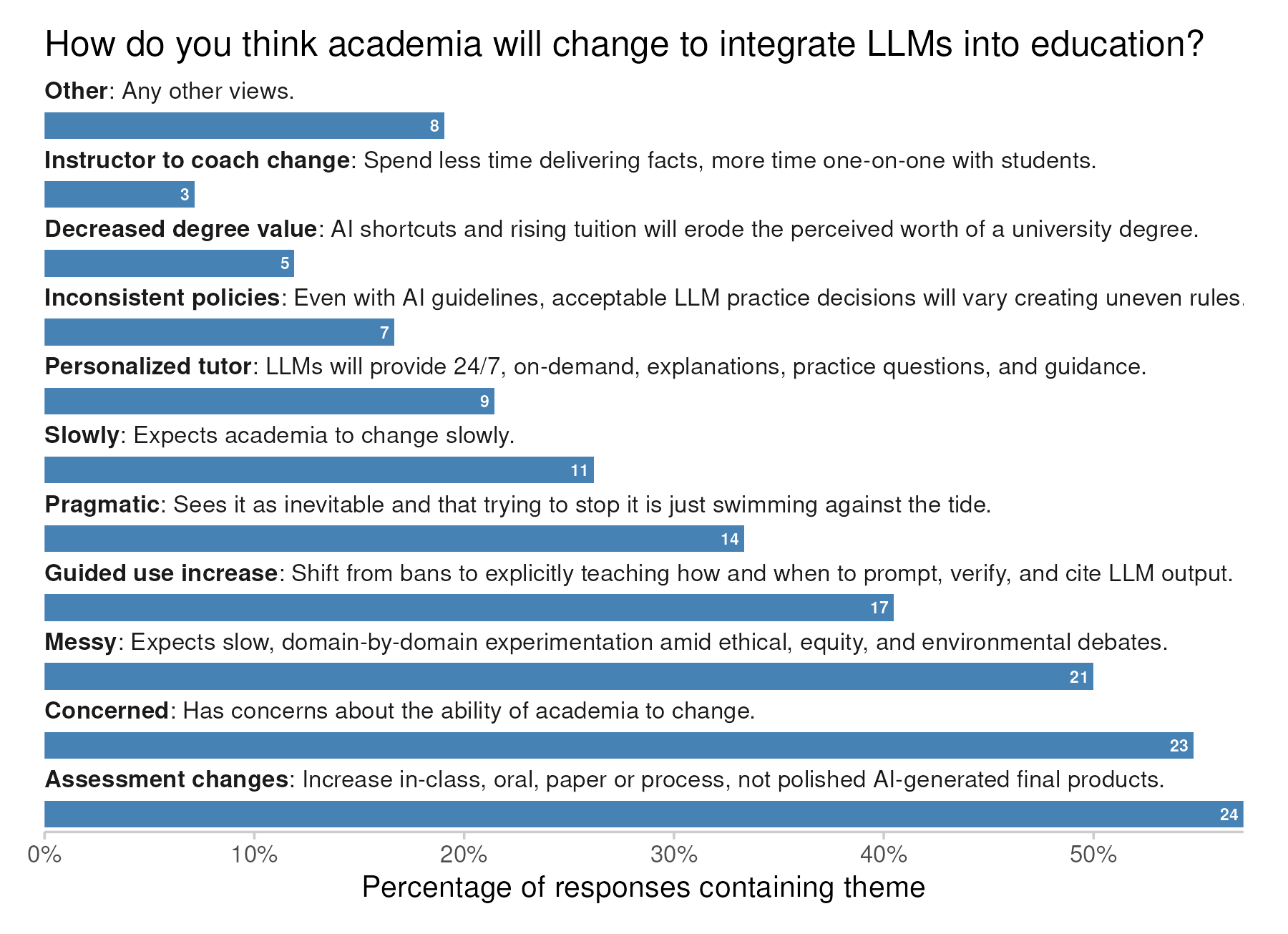}}

}

\caption{\label{fig-academia-change}Most frequent themes in interview
responses to the question ``How do you think academia will change to
integrate LLMs into education?'' The percentage and count of each coded
theme (bolded text above the bar) across the interview responses is
represented by the blue bar and the white number, respectively. Each
interview response could contain more than one theme. A concise
description of the coded theme (unbolded text) follows each coded theme
(bolded text).}

\end{figure}

Finally, respondents were asked ``How do you think academia will change
to integrate LLMs into education?'' (Figure~\ref{fig-academia-change}).
Responses were extensively aligned with ten recurring themes: assessment
changes, concerned, messy, guided use increase, pragmatic, slowly,
personalized tutor, inconsistent policies, decreased degree value, and
instructor to coach change. Many respondents expect assessment changes
(57 per cent), but are concerned about how the change will go (55 per
cent) and think it may be messy (50 per cent).

\begin{quote}
\emph{I don't know. I truly don't know. I think that might be very
university-specific. I think universities, by nature, are just kind of
slow to adapt. Especially now I think everything takes a long time, and
I've also noticed that in large state universities, like the one I'm at,
it's kind of this game of who can tell who what to do, right? So we can
get really generic guidance from higher-level units, but ultimately,
it's departments, and really, it's instructors who are actually
affecting the change.}
\end{quote}

\section{Discussion}\label{sec-discussion}

\subsection{Overview of findings}\label{overview-of-findings}

The purpose of this study was to understand opinions about, and
reactions to, LLMs among data science instructors. We conducted
semi-structured interviews with 42 instructors teaching data science,
and closely related, courses from 33 unique institutions in 10
countries, between 9 June and 9 July 2025. We developed a codebook that
we used to code responses, and then created summary statistics.

The most frequent general view about LLMs was pragmatic with respondents
seeing them as unavoidable and interested in taking advantage of
benefits, while at the same time cautiously balancing costs. When it
came to teaching, the most frequent views were again pragmatic, while
also feeling conflicted especially as to whether LLMs can help with true
learning, and seeing them as likely to be akin to calculators.

Despite this, more than 30 per cent of respondents said they have not
used LLMs to teach students and do not plan to. Most respondents felt
that the goals of education, such as critical thinking and
problem-solving, remained the same now compared with before LLMs were
widely available. As many also felt that students wanted to use LLMs to
finish tasks faster. Finally, more than half of all respondents felt
that academia would change its approach to assessment, such as increased
in-class assessment or the use of oral exams, but many were concerned
about the ability of academia to change, and expected it to be a messy
process.

During the interviews, several respondents shared the ways they have
changed assessments, content, and policies to account for LLM usage.
Many respondents mentioned changing large-scale projects to scaffolded
projects. This meant changing from a single submission of the fully
completed project, to multiple submissions, such as firstly of a plan,
then of an initial analysis, then of an edited draft, and finally of a
completed project. Respondents said this allowed them to see how a
project evolved, also that it mitigated LLM use because they felt using
an LLM to complete all those components would require considerable
effort. Respondents mentioned requiring students to submit a video
alongside their code where they explain the code line by line to
demonstrate conceptual understanding. Multiple respondents referred to
oral assessments as the ideal accurate representation of a student's
understanding. An alternative for larger classrooms could be
course-specific chatbots that were trained on course materials such as
readings, slides, and documents. Instructors would be able to see
anonymized chat records to understand where students were struggling in
their understanding.

\subsection{Responses to LLMs}\label{responses-to-llms}

There are a variety of ways that respondents have changed because of
LLMs. Here we list some of the more interesting strategies which may be
especially of interest to readers. Some respondents partitioned their
LLM use. For instance, one divided the year into four quarters. In the
first quarter, the students were not allowed to use LLMs for anything.
In the second quarter, they were encouraged to use LLMs as a tutor, but
to try to use it for learning itself. In the last half, they were
allowed to use it as their own assistant. Some respondents split
assessments into categories: preparation, application, and secure.
Preparation included online reading quizzes with unlimited attempts and
could potentially be completed with LLM-assistance. Application included
assignments and projects that were done individually and were not meant
to be done with LLM-assistance, but there was no way of enforcing this.
Secure assessments were in the form of an in-person midterm and final,
and LLM-assistance was both not possible and enforced. Every question
was a variation from the reading quiz or programming assignments. The
overall grade in the class was the minimum of the three categories.

Some universities have established computer-based assessment in a
proctored classroom with a computer on lockdown, to allow more authentic
assessment in a controlled environment. For programming courses, some
respondents had added more exercises on explaining and debugging code.
Some respondents had started using programs that require manual
interaction for the questions. Should a student want to use an LLM, it
would require a certain level of engagement to create a prompt. A few
instructors started including more questions that test conceptual
understanding over procedural fluency. One respondent described adding
white-colored writing to their assignment PDF that the LLM would read
and interpret, but a student would not notice. If the student did not
engage with the problem at all, they would not notice that the response
generated by the LLM had been affected by the hidden text. Another
respondent made an assignment where students would have to critically
assess LLM outputs. Rather than the student answering, they would have
to critique an answer. A couple of respondents mentioned having a
greater emphasis on student teaching and student presentations so they
have to take active agency over the material. Finally another respondent
gave students the option of declaring that they had solved their
assignment with AI but by doing so the student would forfeit the option
of getting feedback on their work.

In the ideal case, classrooms would be small. They would be highly
discussion-based, engaged classrooms where learning could be
personalized. But this is not possible for many institutions. While
exams ensure that students are using their own knowledge, there are many
concerns with heavily-weighted exams. They are typically short and only
so much knowledge can be tested. Further, a student may get nervous and
fail to communicate the knowledge they actually have. In lieu of just
having secured assessments, one respondent suggested that instructors
should instead divide their class into two parts: applied and
tools-based learning. In applied, students need to process and
understand information. The assessments are secure and the questions
focus on conceptual understanding. Once the student has a firm
foundation, the instructor can introduce tools-based learning. In
tools-based learning, students can use any tool. Students can use this
opportunity to resourcefully and responsibly use LLMs to make their
learning efficient. This will help them gain the skills to work
efficiently in the workplace. They should constantly be asking
themselves if they can critically vouch for the information the LLM is
supplying. If they cannot, then they are not ready for tools-based
learning.

\subsection{Comparison with existing
literature}\label{comparison-with-existing-literature}

Comparing the recommendations of Ellis and Slade (2023) and Tu et al.
(2024), with what our respondents are actually doing, reveals
considerable overlap. For instance, some respondents are using LLMs to
help draft slides or content, and some respondents are explicitly
teaching how to use LLMs. Some respondents have changed their assessment
in ways that LLMs can help less. In contrast to Ellis and Slade (2023)
and Tu et al. (2024), as LLM capabilities are improving so quickly, few
of our respondents mentioned trying to outsmart them. Instead, they
focused on two tracks of assessment: that which could be completed using
LLMs and those, such as in-person exams or orals, that could not.

Like the recommendation of Ellis and Slade (2023) and Tu et al. (2024),
some respondents have asked students to cite and document LLM usage,
but, in contrast to Ellis and Slade (2023) and Tu et al. (2024), they do
not police compliance or are moving away from this as LLM integration
deepens. While there were some respondents who mentioned ethical
concerns, especially environmental and copyright, there was not an
overwhelming emphasis on adding ethical content to courses by our
respondents, unlike the recommendations of Ellis and Slade (2023) and Tu
et al. (2024).

Many respondents echoed the same concerns of Ellis and Slade (2023) and
Tu et al. (2024) that the lack of foundational knowledge could cause.
Several interviews outlined ways to mitigate such problems in their
assessments. Most respondents agree there is no way to ensure academic
integrity is being followed with unproctored assessments.

\subsection{Limitations}\label{limitations}

There are several weaknesses of our study. The first, is that we focused
on instructors' opinions and do not take into account students'
opinions. There would be considerable benefit from doing this, and there
are plans to do future studies, but it was beyond the scope of this
paper.

Another weakness was the sample size. Although large for qualitative
approaches, our sample is ultimately fairly small. One alternative would
have been to conduct a survey, but we were interested in understanding
the nuance, compromises, and concerns that can be highlighted in
interviews. We feel that interviews allowed respondents to express their
opinions more directly and specifically even if it meant that we have a
smaller sample than would be ideal.

While the broad list of instructors that we contacted covered a variety
of different backgrounds, ultimately we did see some bias in who
responded. This is mostly easily seen in the over-representation of
Canadian respondents. It is possible that results based on a survey,
perhaps conducted by a professional organization, could draw a
respondent pool that is more representative of the broader data science
instructor population.

Instructors were only given a four-week window during summer to respond
and complete the interview. Thus, our sample is biased towards
instructors that had capacity in a constrained window. There were a
number of respondents who said it was a bad time for them, and likely
more who felt the same way but did not respond to the initial email.
There may be something systematically different about those who were
able to respond. That said, even though our sample is biased, there is
value in having concrete examples of instructor opinion and reaction,
and more rapid dissemination of this knowledge will be helpful to the
profession. Future work could perhaps build on ours and use a
larger-scale, more systematic and quantitative, approach.

\subsection{Concluding remarks and future
directions}\label{concluding-remarks-and-future-directions}

When calculators became accessible, instructors noted how it helped
improve efficiency and allowed students to tackle harder problems, but
acknowledged that they could not stop students from using it at home and
worried about students using them without first having established a
foundation of skills (Watters 2015; Ellington 2003). When the internet
became popular, teenagers liked that it could help them with their
schoolwork and many families got internet access because of its
potential as an educational tool, even while acknowledging it enabled
short-cuts for completing assignments (Simon, Graziano, and Lenhart
2001). And as Wikipedia became better known, at the same time as there
being concerns about the accuracy of the information there were calls
for academia to show students how to use it (Wilson 2008). Knight and
Pryke (2012) even found from surveys of academics and students that a
majority used Wikipedia and that students mostly used it in the initial
stages of completing an assessment, while a majority of academics used
Wikipedia independent of whether they allowed students to use it.

Many of the views raised in response to those earlier technologies were
also shared by our respondents in response to LLMs. As such maybe it
will be the case that, similar to how calculators, the internet, and
Wikipedia created a shift in education making it evolve, so too will
LLMs. How will that play out? We do not know, but many instructors are
already changing their assessment and some are changing what they teach.
There is a great deal of anxiety and concern and perhaps in contrast to
calculators, the internet, and Wikipedia, LLM-based tools are being
extensively marketed and pushed both onto instructors and into the tools
that we use.

Our study focused on instructors and we see it as the first part of a
broader examination which should also look at firstly what current
students think of LLMs and secondly what recent graduates are
experiencing in the workforce.

\newpage

\appendix

\section*{Appendix}\label{appendix}
\addcontentsline{toc}{section}{Appendix}

\section{Disclosure Statement}\label{sec-disclosure}

Ana Elisa Lopez-Miranda conducted this work during an internship at the
Investigative Journalism Foundation.

Rohan Alexander and Ana Elisa Lopez-Miranda gratefully acknowledge the
financial support of the Data Sciences Institute at the University of
Toronto, and a NSERC Alliance Grant.

Tiffany Timbers has no financial or non-financial disclosures to share
for this article.

\section{Acknowledgments}\label{sec-acknowledgments}

The authors are grateful to the respondents who so freely gave us their
time.

\section{Data Repository/Code}\label{sec-datacoderepository}

Our code and data are available at:
\url{https://doi.org/10.5281/zenodo.17106979}.

\section{LLM disclosure statement}\label{sec-llmdisclosure}

We used LLMs in the following ways:

\begin{itemize}
\tightlist
\item
  Data coding assistant: Detailed in our methods section. Used to
  provide initial codebook and coding because of the large number of
  interviews.
\item
  Interactive online search tool: Microsoft Copilot, as at July/August
  2025. Used to check for additional relevant literature because of the
  possibility of related work in different disciplines.
\item
  Grammar and spelling assistant: Microsoft Copilot, as at July/August
  2025; OpenAI ChatGPT5 as at September 2025. Used because of the
  difficulty of identifying typos and awkward phrasing.
\item
  Coding assistant: Microsoft Copilot, as at July/August 2025. Used
  because of the efficiency of creating initial code.
\end{itemize}

\section{Email}\label{sec-email}

Dear Professor {[}name{]},

I am wondering if you might have 30 min free sometime in the next couple
of weeks to be interviewed about LLMs and data science education please?
If so, could you please pick a time that suits: {[}add link to
calendly{]}.

By way of background, Ana Elisa Lopez-Miranda, Tiffany Timbers, and I
are interested in understanding how LLMs are being used in data science
courses, how assessment has changed, what you think appropriate use
looks like, and what concerns or opportunities you're thinking about.

Your participation would involve a 30 min one-on-one interview over
Zoom. Nothing you say will be directly attributed to you. The session
will be recorded and we will anonymize the transcript and send it to you
for review. You may fully withdraw at any point up to 10 days after
receiving the anonymized transcript.

We hope to publish our findings in a general interest journal and will
share this with you.

Thank you for your consideration.

Regards,\\
Rohan

\newpage

\section{Interview guide}\label{sec-interview}

\textbf{Introduction}

Hello Professor,

How are you doing today?

Before we begin, is it alright if I record this meeting?

It's great to finally meet you. My name is Ana Elisa Lopez-Miranda. I'm
a fourth year student at the University of Toronto Mississauga and I'll
be conducting the interview. Thank you so much for taking the time to do
this. This meeting will take 30 minutes.

(If the interviewee has not signed the consent form sent earlier that
day yet, give them 5 minutes to do so.)

Is there anything I can clarify regarding the consent form?

First, I will give an overview of the project before going into the 14
questions I have prepared and wrapping up shortly before \_\_.

This project seeks to explore how Large Language Models (LLMs) are being
used by faculty in data science courses. We will investigate how faculty
have changed their assessments to account for the existence of LLMs,
faculty opinions about the appropriate use of LLMs in teaching data
science, faculty reasons for using or not using LLMs in their teaching
and the potential advantages and disadvantages of each approach.

There are no right or wrong answers.

\textbf{Questions}

{[}Not all questions were asked in every interview.{]}

\begin{enumerate}
\def\labelenumi{\arabic{enumi}.}
\tightlist
\item
  Will you please state your name, and academic background.
\item
  How long have you been teaching?
\item
  Where and what do you normally teach?
\item
  What is the average size of your classes and the typical level
  (u/g/grad)?
\item
  What are your general views on LLMs?
\item
  What are your views on using LLMs to teach students?
\item
  Have you used LLMs to teach students, if so, how have you used LLMs to
  teach students?
\item
  What LLM interface do you use to teach students (i.e.~ChatGPT, Claude,
  Gemini, etc.)?
\item
  Have you noticed a change in grade distributions pre-AI to now?
\item
  Do you think there's a difference in the goal of education now vs
  pre-AI?
\item
  Do you find there's a difference in students' attitudes and motivation
  towards learning now?
\item
  What are the benefits of using LLMs to teach students?
\item
  What concerns do you have of using LLMs to teach students?
\item
  In the past few years, how have you changed your assessment given the
  ubiquity of LLMs?
\item
  In the coming few years, how do you think you will change your
  assessment?
\item
  How do you create meaningful take-home assignments and projects that
  are not AI agnostic, but are not devaluing the learning experience for
  the students?
\item
  Do you ask your students to cite LLMs in their work?
\item
  Have you seen instances of hallucinations or other made-up aspects in
  submitted work?
\item
  Putting hallucinations to one side, do you think you can tell when
  some code or writing has been produced by LLMs?
\item
  How do you think academia will change to integrate LLMs into
  education?
\item
  Now that you know what aspects I'm interested in, is there anything
  else that you think I should be asking?
\item
  How do you ensure your learning objectives are being learnt?
\item
  Is there anyone that you recommend we contact about this project?
\end{enumerate}

\textbf{Conclusion}

Those are all the questions. Before we conclude, I would like to review
that you may withdraw from the study any time up to 10 days after we
have sent you the anonymized transcript.

Do you have any questions for me?

Thank you very much for your time. I hope you have a great rest of your
day.

\newpage

\section{Anonymization}\label{sec-anonymization}

The transcripts were anonymized by:

\begin{itemize}
\tightlist
\item
  Removing the name of each person.
\item
  Removing the institution.
\item
  Generalizing or removing program and course names.
\item
  Generalizing or removing aspects of courses that could be identifying.
\item
  Removing city and country references.
\item
  Removing details of the people the respondents suggest we contact.
\end{itemize}

\newpage

\section{Codebook}\label{sec-codebook}

\begin{itemize}
\tightlist
\item
  Where are you based?

  \begin{itemize}
  \tightlist
  \item
    Canada
  \item
    US
  \item
    Other
  \end{itemize}
\item
  Will you please state your academic background? {[}This is typically
  their PhD, otherwise their highest degree, or post doc.{]}

  \begin{itemize}
  \tightlist
  \item
    Biostatistics
  \item
    Computer Science
  \item
    Statistics
  \item
    Other
  \end{itemize}
\item
  How long have you been teaching?

  \begin{itemize}
  \tightlist
  \item
    1--5 years
  \item
    6--10 years
  \item
    11--20 years
  \item
    20+ years
  \end{itemize}
\item
  What do you normally teach?

  \begin{itemize}
  \tightlist
  \item
    Computer Science
  \item
    Data Science
  \item
    Statistics
  \item
    Other
  \end{itemize}
\item
  What is the average size of your classes? {[}This was sometimes
  difficult to work out because the respondent taught multiple classes.
  Tried to get a sense of which one they had in mind when answering the
  rest of the questions.{]}

  \begin{itemize}
  \tightlist
  \item
    1--10 students
  \item
    11--20 students
  \item
    21--50 students
  \item
    51--100 students
  \item
    101--200 students
  \item
    201--500 students
  \item
    500+ students
  \end{itemize}
\item
  What is the typical level?

  \begin{itemize}
  \tightlist
  \item
    Undergraduate
  \item
    Graduate
  \end{itemize}
\item
  What are your general views on LLMs?

  \begin{itemize}
  \tightlist
  \item
    Ethical concerns: Concerns about LLMs on moral, ownership, bias, or
    sustainability grounds.
  \item
    LLM capability concerns: Concerns about the difference between
    advertised and current capability; may worry about unpredictable
    long-term effects.
  \item
    Capacity concerns: Worried that academic institutions and
    instructors may lack the skills or bandwidth to keep up.
  \item
    Learning integrity concerns: Worries that students can use LLMs to
    avoid thinking or deeper engagement with content.
  \item
    Pragmatic: Thinks that LLMs are unavoidable and interested in taking
    advantage of benefits, cautiously, and balanced with costs.
  \item
    Specific uses: Uses LLMs tactically for drafting, debugging, or idea
    generation.
  \item
    Excitement: Excited about redesigning courses to teach with, and
    about, LLMs; thinks LLM fluency is essential; excited about new
    technology.
  \item
    Other: Any other views.
  \end{itemize}
\item
  What are your views on using LLMs to teach students?

  \begin{itemize}
  \tightlist
  \item
    Conflicted: Has experimented with LLMs but has concerns about
    whether they can help with true learning.
  \item
    Student engagement: Co-creates class rules about LLM use.
  \item
    Pragmatic: Allows LLMs because policing is impossible; stresses
    fairness and citation; sees it as inevitable and that trying to stop
    it is just swimming against the tide.
  \item
    Personalized tutor: Sees benefit of LLMs providing 24/7, on-demand,
    explanations, examples, practice questions, and guidance.
  \item
    Calculators: Thinks of LLMs as the next everyday tool, similar to
    calculators or Wikipedia.
  \item
    Opposed: Opposes LLMs and wants to keep them out of teaching.
  \item
    Unsure: Is waiting to adopt them until more is known.
  \item
    Concerned: Is worried that LLMs make avoiding work easier and
    undermines known, but difficult, ways of learning.
  \item
    Foundations: Thinks LLMs should not be allowed until students master
    core skills.
  \item
    Other: Any other views.
  \end{itemize}
\item
  Have you used LLMs to teach students, if so, how have you used LLMs to
  teach students?

  \begin{itemize}
  \tightlist
  \item
    None: No classroom use and does not plan to.
  \item
    Planned: No classroom use yet, but plans to.
  \item
    Prep only: Uses LLMs to draft materials like rubrics, exercises, and
    slides, but students do not interact with LLMs.
  \item
    Demo use: Occasionally gives in-class demonstrations of an LLM
    feature; student use optional and brief.
  \item
    Guided use: Provides structured workshop or guided activities that
    expect students to use LLMs responsibly.
  \item
    Extensive: LLMs are woven into many course tasks, such as chatbots,
    and there is regular student interaction.
  \item
    Other: Any other views.
  \end{itemize}
\item
  What LLM interface do you use to teach students (i.e.~ChatGPT, Claude,
  Gemini, etc.)?

  \begin{itemize}
  \tightlist
  \item
    OpenAI: Likely ChatGPT but could be other models also.
  \item
    Anthropic: Likely Claude but could be other models also.
  \item
    Google: Likely Gemini but could be other models also.
  \item
    GitHub: Must be Copilot.
  \item
    Other
  \end{itemize}
\item
  Have you noticed a change in grade distributions pre-AI to now?

  \begin{itemize}
  \tightlist
  \item
    Inflated: Thinks grades are noticeably higher on tasks that could be
    done by an LLM.
  \item
    Stable: No clear shift in grade patterns since LLMs became widely
    available.
  \item
    Bimodal: Has seen a wider gap, with more top and low scores, fewer
    in the middle.
  \item
    Compressed: Grades have bunched toward the middle; fewer extreme
    good or extreme bad.
  \item
    Confounded: Thinks change cannot be isolated to LLMs because of the
    pandemic, new rubrics, or other factors.
  \item
    Office hours: Has seen office hours attendance reduced.
  \item
    Other: Any other views.
  \end{itemize}
\item
  Do you think there's a difference in the goal of education now vs
  pre-AI?

  \begin{itemize}
  \tightlist
  \item
    Unchanged: Thinks the main aims of education, such as critical
    thinking and problem-solving, are the same.
  \item
    Conceptual change: See a change from students needing to know rote
    skills to them focusing on higher-level understanding because
    low-level tasks can be done by LLMs.
  \item
    AI-literacy: The goal of education now includes teaching students to
    use, and critique, LLMs.
  \item
    Assessment redesign: Thinks LLMs mean assignments and tests need to
    be re-written to account for potential LLM use.
  \item
    Uncertain: Thinks the impact is still unclear.
  \item
    Rethinking: Believes education was misaligned and that LLMs are
    forcing a fundamental rethinking.
  \item
    Credentialling: Questions the long-term value of university degrees
    if LLMs can do all the same tasks.
  \item
    Other: Any other views.
  \end{itemize}
\item
  Do you find there's a difference in students' attitudes and motivation
  towards learning now?

  \begin{itemize}
  \tightlist
  \item
    Same: Motivation is much the same; LLMs has not changed it.
  \item
    Speed: Students want to use LLMs to finish tasks faster; increased
    get it done quickly attitude; product valued over process.
  \item
    Bimodal: Bigger divide between a highly driven group and a
    disengaged AI-dependent group.
  \item
    Confounded: Changes are hard to identify because AI impacts are
    mixed with other disruptions, such as the pandemic.
  \item
    Motivation concerns: Sees student expressing uncertainty of gaining
    skills that AI can do.
  \item
    Other: Any other views.
  \end{itemize}
\item
  What are the benefits of using LLMs to teach students?

  \begin{itemize}
  \tightlist
  \item
    Skills: Gives students skills with a tool that is, or will, be in
    the workplace, and provides an opportunity for students to
    demonstrate appropriate and ethical use.
  \item
    Engagement: Hype around AI draws attention and increases class
    participation.
  \item
    Personalized tutor: LLMs provide 24/7, on-demand, explanations,
    examples, practice questions, and guidance.
  \item
    Efficiency: Makes drafting, editing, coding, and prep such as
    practice questions and slides, faster, for both students and
    instructors.
  \item
    Scaling: Can automate individualized comments, rubrics, and
    formative checks in large classes.
  \item
    Other: Any other views.
  \end{itemize}
\item
  What concerns do you have of using LLMs to teach students?

  \begin{itemize}
  \tightlist
  \item
    De-skilling: Students skip practice questions and exercises, and may
    never build core mental models.
  \item
    Illusory mastery: AI-generated work can hide gaps, inflate grades,
    and result in misplaced confidence that is revealed in exams.
  \item
    Black-boxes: Hallucinations, bias, and unverifiable answers can
    mislead learners.
  \item
    Shortcuts: LLMs allow easy cheating and lowers the effort required
    of students; it can devalue authentic learning.
  \item
    Equity: Paywalled or premium models may advantage wealthier
    students.
  \item
    Moving target: AI can change faster than courses and university
    policies can adapt.
  \item
    Other: Any other views.
  \end{itemize}
\item
  In the past few years, how have you changed your assessment given the
  ubiquity of LLMs?

  \begin{itemize}
  \tightlist
  \item
    In-class return: Shift back to in-class paper/computer exams/quizzes
    and orals.
  \item
    Reduced take-home weight: Reduce the weight of work done outside
    controlled settings.
  \item
    Process proof: Demand version histories, drafts, or commit history.
  \item
    AI-integrated tasks: Build prompts or critique-the-LLM activities
    directly into assignments.
  \item
    Authenticity reduction: Has changed assessment in a way that reduces
    the authenticity.
  \item
    No change: No changes implemented.
  \item
    Other: Any other views.
  \end{itemize}
\item
  In the coming few years, how do you think you will change your
  assessment?

  \begin{itemize}
  \tightlist
  \item
    In-class return: Shift back to in-class paper/computer exams/quizzes
    and orals.
  \item
    Reduced take-home weight: Reduce the weight of work down outside
    controlled settings.
  \item
    Process proof: Aspects such as version history, drafts, or commit
    history are required.
  \item
    AI-integrated tasks: Build prompts, or critique-the-LLM, activities
    directly into assignments.
  \item
    Unsure: Finds it hard to tell how it will change.
  \item
    No change: No changes planned.
  \item
    Other: Any other views.
  \end{itemize}
\item
  Do you ask your students to cite LLMs in their work?

  \begin{itemize}
  \tightlist
  \item
    No: Makes no request for LLM disclosure. Views LLM citation as
    inappropriate, impractical or meaningless.
  \item
    Guided: Gives examples of when to cite versus not to train judgment.
  \item
    Honor: Asks to cite and document, but does not police compliance.
  \item
    Yes: Citation of AI-use statement is compulsory and graded.
  \item
    Other: Any other views.
  \end{itemize}
\item
  Have you seen instances of hallucinations or other made-up aspects in
  submitted work?

  \begin{itemize}
  \tightlist
  \item
    Obvious fabrication: Submitted work includes invented formulas, data
    sets, or citations.
  \item
    Stylistic tells: Hollow prose, over-commented code, or phrasing
    common to LLMs, suggests LLM use.
  \item
    Self-corrected: Believes that if students find hallucinations while
    working they fix them before handing in work; especially relevant
    for coding.
  \item
    Undetected or unsure: Instructor rarely spots hallucinations, or
    cannot verify AI origin, or does not believe it is possible to tell.
  \item
    No: Instructor does not believe they have seen hallucinations.
  \item
    Other: Any other views.
  \end{itemize}
\item
  Putting hallucinations to one side, do you think you can tell when
  some code or writing has been produced by LLMs?

  \begin{itemize}
  \tightlist
  \item
    Obvious fabrication: Submitted work includes invented formulas, data
    sets, or citations.
  \item
    Pattern spotting: Detects AI through tone, over-commented code, odd
    functions never taught, or other code smells.
  \item
    Identical submissions: Identical wording or variable names across
    many students suggests LLM use.
  \item
    Asked for explanation: Instructor confirms suspicion by having the
    student verbally walk through the code or text.
  \item
    Undetected or unsure: Believes detection accuracy is difficult or
    impossible, and often unverifiable.
  \item
    Pointless: Does not believe that LLM use matters.
  \item
    No: Instructor does not believe it is possible to tell.
  \item
    Other: Any other views.
  \end{itemize}
\item
  How do you think academia will change to integrate LLMs into
  education?

  \begin{itemize}
  \tightlist
  \item
    Guided use increase: Thinks academia will shift from bans to
    explicitly teaching how and when to prompt, verify, and cite LLM
    output.
  \item
    Instructor to coach change: Thinks instructors will spend less time
    delivering facts, more time mentoring, troubleshooting and
    one-on-one with students.
  \item
    Assessment changes: Expects an increase in oral, in-class, paper or
    version-tracked tasks that test process, not polished AI-generated
    final products.
  \item
    Inconsistent policies: Expects that even when task-forces draft AI
    guidelines, acceptable LLM practice decisions stays at a department,
    or instructor-specific, basis, creating uneven rules.
  \item
    Personalized tutor: Expects that LLMs will provide 24/7, on-demand,
    explanations, examples, practice questions, and guidance. These may
    replace entry-level courses and enable large courses to scale.
  \item
    Decreased degree value: Worries that AI shortcuts and rising tuition
    will erode the perceived worth of a university degree.
  \item
    Messy: Expects slow, domain-by-domain experimentation amid ethical,
    equity, and environmental debates.
  \item
    Concerned: Has concerns about the ability of academia to change.
  \item
    Slowly: Expects academia to change only slowly.
  \item
    Pragmatic: Sees it as inevitable and that trying to stop it is just
    swimming against the tide.
  \item
    Other: Any other views.
  \end{itemize}
\end{itemize}

\newpage

\section*{References}\label{references}
\addcontentsline{toc}{section}{References}

\phantomsection\label{refs}

\end{document}